\documentclass[bibyear]{aa}  

\usepackage{graphicx}
\usepackage{txfonts}
\usepackage{dsfont}
\usepackage{hyperref}

\usepackage{subcaption}

\usepackage{pdflscape}
\usepackage{tikz}
\usetikzlibrary{arrows,arrows.meta,decorations.pathmorphing,decorations.markings,backgrounds,positioning,fit,calc,shapes,shapes,matrix}
\usepackage{varwidth}
\usepackage{enumitem}

\usepackage{natbib}

\bibpunct{(}{)}{;}{a}{}{,} % to follow the A&A style

\newcommand{\nifty}{\textsc{NIFTy}}
\newcommand{\niftyo}{\textsc{NIFTy\,1}}
\newcommand{\niftyt}{\textsc{NIFTy\,3}}

\newcommand{\iu}{{i\mkern1mu}}

\newcommand{\red}[1]{{\color[rgb]{1,0,0} #1}}

\usepackage{xcolor}
\usepackage{verbatim}

\definecolor{green}{rgb}{0,1,00}
\definecolor{lightgreen}{rgb}{0,0.5,0}
\definecolor{red}{rgb}{1,0,0}
\definecolor{lightred}{rgb}{0.5,0,0}

\newcounter{inCounter}[section]
\newcommand{\iin}{\refstepcounter{inCounter}{\color{lightgreen}In [{\color{green}\arabic{inCounter}}]:}}
\newcommand{\out}{{\color{lightred}Out[{\color{red}\arabic{inCounter}}]:}}

\definecolor{maroon}{cmyk}{0, 0.87, 0.68, 0.32}
\definecolor{halfgray}{gray}{0.55}
\definecolor{ipython_frame}{RGB}{207, 207, 207}
\definecolor{ipython_bg}{RGB}{247, 247, 247}
\definecolor{ipython_red}{RGB}{186, 33, 33}
\definecolor{ipython_green}{RGB}{0, 128, 0}
\definecolor{ipython_cyan}{RGB}{64, 128, 128}
\definecolor{ipython_purple}{RGB}{170, 34, 255}

\usepackage{listings}
\lstdefinelanguage{iPython}{
	morekeywords={access,and,break,class,continue,def,del,elif,else,except,exec,finally,for,from,global,if,import,in,is,lambda,not,or,pass,print,raise,return,try,while},%
	morekeywords=[2]{abs,all,any,basestring,bin,bool,bytearray,callable,chr,classmethod,cmp,compile,complex,delattr,dict,dir,divmod,enumerate,eval,execfile,file,filter,float,format,frozenset,getattr,globals,hasattr,hash,help,hex,id,input,int,isinstance,issubclass,iter,len,list,locals,long,map,max,memoryview,min,next,object,oct,open,ord,pow,property,range,raw_input,reduce,reload,repr,reversed,round,set,setattr,slice,sorted,staticmethod,str,sum,super,tuple,type,unichr,unicode,vars,xrange,zip,apply,buffer,coerce,intern},%
	sensitive=true,%
	morecomment=[l]\#,%
	morestring=[b]',%
	morestring=[b]",%
	morestring=[s]{'''}{'''},% used for documentation text (mulitiline strings)
	morestring=[s]{"""}{"""},% added by Philipp Matthias Hahn
	morestring=[s]{r'}{'},% `raw' strings
	morestring=[s]{r"}{"},%
	morestring=[s]{r'''}{'''},%
	morestring=[s]{r"""}{"""},%
	morestring=[s]{u'}{'},% unicode strings
	morestring=[s]{u"}{"},%
	morestring=[s]{u'''}{'''},%
	morestring=[s]{u"""}{"""},%
	literate=
	{á}{{\'a}}1 {é}{{\'e}}1 {í}{{\'i}}1 {ó}{{\'o}}1 {ú}{{\'u}}1
	{Á}{{\'A}}1 {É}{{\'E}}1 {Í}{{\'I}}1 {Ó}{{\'O}}1 {Ú}{{\'U}}1
	{à}{{\`a}}1 {è}{{\`e}}1 {ì}{{\`i}}1 {ò}{{\`o}}1 {ù}{{\`u}}1
	{À}{{\`A}}1 {È}{{\'E}}1 {Ì}{{\`I}}1 {Ò}{{\`O}}1 {Ù}{{\`U}}1
	{ä}{{\"a}}1 {ë}{{\"e}}1 {ï}{{\"i}}1 {ö}{{\"o}}1 {ü}{{\"u}}1
	{Ä}{{\"A}}1 {Ë}{{\"E}}1 {Ï}{{\"I}}1 {Ö}{{\"O}}1 {Ü}{{\"U}}1
	{â}{{\^a}}1 {ê}{{\^e}}1 {î}{{\^i}}1 {ô}{{\^o}}1 {û}{{\^u}}1
	{Â}{{\^A}}1 {Ê}{{\^E}}1 {Î}{{\^I}}1 {Ô}{{\^O}}1 {Û}{{\^U}}1
	{œ}{{\oe}}1 {Œ}{{\OE}}1 {æ}{{\ae}}1 {Æ}{{\AE}}1 {ß}{{\ss}}1
	{ç}{{\c c}}1 {Ç}{{\c C}}1 {ø}{{\o}}1 {å}{{\r a}}1 {Å}{{\r A}}1
	{€}{{\EUR}}1 {£}{{\pounds}}1,
	literate=
	*{+}{{{\color{ipython_purple}+}}}1
	{-}{{{\color{ipython_purple}-}}}1
	{*}{{{\color{ipython_purple}$^\ast$}}}1
	{/}{{{\color{ipython_purple}/}}}1
	{^}{{{\color{ipython_purple}\^{}}}}1
	{?}{{{\color{ipython_purple}?}}}1
	{!}{{{\color{ipython_purple}!}}}1
	{\%}{{{\color{ipython_purple}\%}}}1
	{|}{{{\color{ipython_purple}|}}}1
	{\&}{{{\color{ipython_purple}\&}}}1
	{~}{{{\color{ipython_purple}~}}}1
	{==}{{{\color{ipython_purple}==}}}2
	{<=}{{{\color{ipython_purple}<=}}}2
	{>=}{{{\color{ipython_purple}>=}}}2
	{np.float}{{{np.float}}}7
	{np.complex}{{{np.complex}}}9
	{+=}{{{+=}}}2
	{-=}{{{-=}}}2
	{*=}{{{$^\ast$=}}}2
	{/=}{{{/=}}}2,
	commentstyle=\color{ipython_cyan}\ttfamily,
	stringstyle=\color{ipython_red}\ttfamily,
	keepspaces=true,
	showspaces=false,
	showstringspaces=false,
	rulecolor=\color{black},
	frame=l,
	frameround={t}{t}{t}{t},
	framexleftmargin=0mm,
	numbers=left,
	numberstyle=\tiny\color{black},
	basicstyle=\small\ttfamily,
	keywordstyle=\color{ipython_green}\ttfamily,
	xleftmargin=18pt,
	escapechar=|,escapebegin=\color{ipython_green},
	gobble=4,
	tabsize=4,
}

\begin{document}

   \title{NIFTy 3 -- Numerical Information Field Theory}

   \subtitle{A Python framework for multicomponent signal inference on HPC clusters}

   \author{Theo Steininger \inst{1,2}
           \and
           Jait Dixit \inst{3}
           \and
           Philipp Frank \inst{1,2}
           \and
           Maksim Greiner \inst{1,2}
           \and
           Sebastian Hutschenreuter \inst{1,2}
           \and
           Jakob Knollmüller \inst{1,2}
           \and
           Reimar Leike \inst{1,2}
           \and
           Natalia Porqueres \inst{1,2}
           \and
           Daniel Pumpe \inst{1,2}
           \and
           Martin Reinecke \inst{1}
           \and
           Matevž Šraml \inst{1,2}
           \and
           Csongor Varady \inst{3}
           \and
           Torsten En{\ss}lin \inst{1,2}
           }

   \institute{Max Planck Institute for Astrophysics, 
              Karl-Schwarzschild-Str. 1, 
              D-85741 Garching, Germany
         \and
	         Ludwig-Maximilians-Universität München
	         Geschwister-Scholl-Platz 1,
	         80539 Munich, 
	         Germany         
         \and
             Technische Universität München,
             Arcisstr. 21,
             80333 München, Germany
             }

   \date{Received August 03, 2017}

% \abstract{}{}{}{}{} 
% 5 {} token are mandatory
 
\abstract{
\nifty, ``Numerical Information Field Theory'', is a software framework designed to ease the development and implementation of field inference algorithms.
Field equations are formulated independently of the underlying spatial geometry allowing the user to focus on the algorithmic design. 
Under the hood, \nifty\ ensures that the discretization of the implemented equations is consistent. 
This enables the user to prototype an algorithm rapidly in 1D and then apply it to high-dimensional real-world problems.
This paper introduces \niftyt, a major upgrade to the original \nifty\ framework. 
\niftyt\ allows the user to run inference algorithms on massively parallel high performance computing clusters without changing the implementation of the field equations. 
It supports n-dimensional Cartesian spaces, spherical spaces, power spaces, and product spaces as well as transforms to their harmonic counterparts. 
Furthermore, \niftyt\ is able to treat non-scalar fields.
The functionality and performance of the software package is demonstrated with example code, which implements a real inference algorithm from the realm of information field theory.
\niftyt\ is open-source software available under the GNU General Public License v3 (GPL-3) at \url{https://gitlab.mpcdf.mpg.de/ift/NIFTy/}. 
}


   \maketitle
%
%-------------------------------------------------------------------

\section{Introduction}
Physical quantities are only accessible to the observer via measurements. 
Those measurements are never a perfect mapping of the quantity of interest, the so-called signal. 
They are virtually always subject to noise, which cannot be distinguished from the signal.
Usually measurements capture the quantity of interest only indirectly, be it by event rates, receiver voltages, absorption or reflection patterns, or photon counts. 
Such indirect measurements cannot capture all properties of the signal.\\
There are two ends from which this shortcoming can be approached. 
The first one is to improve the measurement itself. 
This includes the development of more precise instruments, better observation strategies, and completely new experimental designs. 
The second one is to improve the processing of the data, i.e.\ the way how the signal is estimated once the data have been taken. 
This is the domain of information theory.\\
Modern information theory is based on the work of \citet{doi:10.1119/1.1990764}, \citet{BLTJ:BLTJ1338} and \cite{1949wiener}. 
It has seen tremendous advances since the development of computers, which allow the calculation of statistical estimates in the absence of closed form solutions through methods such as gradient descent \citep[e.g.][]{FlPo63a} and sampling \citep[e.g.][]{doi:10.1063/1.1699114}. 
In the last decades, data sets have become larger and larger, which makes their processing feasible only on a computer, even if a closed form solution is available.\\
Many physical signals are fields, quantities that are described as functions over a continuous domain (e.g.\ time or space). 
Applying the rules of information theory to fields means applying the calculus of probabilities to functional spaces. 
This leads to a statistical field theory, which we call information field theory (IFT) \citep{2009PhRvD..80j5005E, 2010PhRvE..82e1112E}. 
Since fields have an infinite amount of degrees of freedom by nature, the calculations cannot directly be implemented on a computer, which can only store and process a finite amount of bits. 
Careful considerations have to be made in discretizing the continuous equations to calculate statistical estimates that are independent of the chosen resolution, as long as all scales that are imprinted in the data are resolved. 
These considerations led to the development of \nifty\ \citep{2013A&A...554A..26S}.\\
\nifty\ -- ``Numerical Information Field Theory'' -- is a software framework in which algorithms for information extraction from data and signal reconstruction are implemented. 
Its purpose is to ease the technical as well as conceptual difficulties which arise when working with IFT. 
To that end \nifty\ was designed in a way that equations involving fields are formulated independently of the underlying spatial geometry and discretization. 
Thus, an algorithm implemented with \nifty\ to infer a field living on a sphere, can be transformed to infer a field living in a three-dimensional Cartesian space by the change of a single line of code.
\nifty\ is a development framework written in \textsc{Python}\footnote{\textsc{Python} homepage \url{http://www.python.org/}}. 
The structure \nifty\ brings in terms of a development framework is bundled with the power of external compiled-language modules for numerical efficiency. \\
As it greatly reduced the complexity of implementing algorithms for field inference, many applications of IFT have been implemented using \nifty,
%Hier vielleicht schon eine kurze Liste mit den wichtigsten (5?) Projekten reintun, am besten die Projekte mit echten Daten 
 most notably the reconstruction of the primordial scalar potential \citep{2015JCAP...02..041D}, the estimation of extragalactic Faraday rotation \citep{2015A&A...575A.118O}, and the denoised, deconvolved, and decomposed Fermi $\gamma$-ray sky \citep{2015A&A...581A.126S}.
% hier noch eine lange liste mit fast allen nifty anwendunguen bis 2016
%\citep[e.g.,][]{2013PhRvE..87c2136O, 2013PhRvD..88j3516D, 2014IAUS..306...16O, 2014JCAP...06..048D, 2014PhRvE..90d3301E, 2015PhRvE..91a3311D, 2015A&A...574A..74S, 2015JCAP...02..041D, 2015A&A...575A.118O, 2015PhRvE..92a3302D, 2015A&A...581A..59J, 2015A&A...581A.126S, 2016A&A...586A..76J, 2016A&A...590A..59G, 2016arXiv160504317G, 2016A&A...591A..13V, 2016PhRvE..94a2132P}. %vielleicht etwas zusammenstreichen :-)
However, more and more complex and ambitious field inference projects brought the original \nifty\ package to its limits. 
To advance further, the capability to make use of massively parallel high performance computing clusters, 
	treat products of spaces,
	and work with non-scalar fields is needed. 
These requirements have led to a complete redesign of \nifty. 
This new software framework, \niftyt, which additionally includes also numerous minor advances, is presented here and is available at \url{https://gitlab.mpcdf.mpg.de/ift/NIFTy}. 

\section{Problem Description}\label{sec:problem_description}
\subsection{Information Field Theory}\label{sec:ift}
In general, a signal inference problem tries to invert a measurement equation 
\begin{equation}\label{eq:measurement_equation}
d = f(s, n), 
\end{equation}
that describes how the obtained data $d$ depends on the unknown 
signal $s$ and further nuisance parameters $n$. 
Actually, we want to know $s$ and are not interested in $n$, but the latter also influences the measurement outcome. 
The problem is that the function $f$ is not necessarily invertible in $s$, and that the nuisance parameters may not be known.\\
This problem is particularly severe when the signal is a field, viz.\ a continuous function $s(x)$ over a manifold $\Omega$. 
In this case the number of unknowns is infinite, whereas the number of knowns, the components of the data vector, is finite. 
The reconstruction of a field from finite data therefore always requires the usage of additional information. 
Optimally this is given in the form of a joint probability distribution $\mathcal{P}(s, n)$ for all the unknowns.
Hence, this induces a joint probability for data and signal,
\begin{eqnarray}
\mathcal{P}(d,s)
&=& \int \mathcal{D}n\, \mathcal{P}(d| s, n)\, \mathcal{P}(s, n)\nonumber\\
&=& \int \mathcal{D}n\, \delta(d-f(s, n))\, \mathcal{P}(s, n),
\end{eqnarray}
where $\int \mathcal{D}n$ denotes the integration over all degrees of freedom of $n$.
The posterior distribution for the unknown signal is then given by Bayes' theorem:
\begin{equation}
	\mathcal{P}(s|d)
	=\frac{\mathcal{P}(d,s)}{\mathcal{P}(d)}
	=\frac{e^{-\mathcal{H}(d,s)}}{\mathcal{Z}(d)},
\end{equation}
where we introduced the language of statistical field theory by defining
\begin{eqnarray}
	\mathcal{H}(d,s) &:=& -\ln \mathcal{P}(d,s) \quad \mathrm{and}\\
	\mathcal{Z}(d)   &:=& \mathcal{P}(d) = \int \mathcal{D}s\, e^{-\mathcal{H}(d,s)}.
\end{eqnarray}
Hereafter we focus on a special case that leads to the well-known Wiener filter theory.

\subsection{Wiener Filter Theory}\label{sec:WF}

A simple example for an inference problem is given by a linear measurement of a signal $s$ with additive and signal independent noise $n$
\begin{equation}\label{eq:data-equation}
d = Rs + n
\end{equation}
of a Gaussian signal 
\begin{equation}
	s \hookleftarrow\mathcal{P}(s)=\mathcal{G}(s,S)=\frac{1}{|2\pi S|^\frac{1}{2}}\exp{\left( -\frac{1}{2}\, s^\dagger S^{-1}s\right)}
\end{equation}
with additive and signal independent Gaussian noise $n\hookleftarrow\mathcal{P}(n| s)=\mathcal{G}(n,N)$ and known signal and noise covariances $S=\langle s\,s^\dagger \rangle_{({s})}$ and $N=\langle n\,n^\dagger \rangle_{(n)}$. $\dagger$ denotes transposition and complex conjugation so that $s^\dagger t=\int_{\Omega}dx\, s^*(x)\,t(x)$ is a scalar and $\langle f(x) \rangle_{(x|y)}=\int \mathcal{D}x\,\mathcal{P}(x|y)\,f(x)$ is the expectation of $f(x)$ over $\mathcal{P}(x|y)$.\\
The response $R$ maps the continuous field $s$ onto a discrete data vector $d$. 
For individual entries of this data vector equation \ref{eq:data-equation} reads
\begin{equation}\label{eq:data-equation-discretized}
d_i = \int_{\Omega}\mathrm{d}x \; R_i(x)s(x) + n_i
\end{equation}
Here, $i \in \{1,\dots,N_\mathrm{data}\}$ with $N_\mathrm{data} \in \mathds{N}$ is the index of the data vector's entries. 
$\Omega$ is the physical manifold the signal is defined on. \\
Since equation \ref{eq:data-equation} maps a continuous space with infinitely many degrees of freedom onto a finite-dimensional data vector and involves an additive stochastic noise term, the problem is not uniquely invertible.
This means that many combinations of $s$- and $n$-realizations can result in the same data vector.
One can now apply an inference algorithm to find the `best'\footnote{The exact meaning of `best' depends on the algorithm.} estimate for $s$.  One possibility here is the Wiener filter, which provides the posterior mean signal as the field estimate,
\begin{equation}
 m = \left( S^{-1} + R^\dagger N^{-1} R \right)^{-1} R^\dagger N^{-1} d,
 \label{eq:WF_condensed}
\end{equation}
where $S$ is the covariance of the signal and $N$ the covariance of the noise. The superscript $\dagger$ denotes complex conjugation and transposition.
Spelled out, this means the equation
\begin{equation}
 \int_{\Omega}\mathrm{d}y \left( S^{-1}(x, y) + \sum\limits_{i,j=1}^{N_{\mathrm{data}}} R_i(x) N^{-1}_{ij} R_j(y) \right) m(y) = \sum\limits_{i,j=1}^N R_i(x) N^{-1}_{ij} d_j
 \label{eq:WF_spelled_out}
\end{equation}
is solved for the estimate $m(x)$.\\
Equation \ref{eq:data-equation-discretized} already shows the three main components of data equations in the area of signal reconstruction: 
	\emph{operators} ($R$) that act on \emph{fields} ($s$) which are defined on a \emph{space} ($\Omega$). 
The signal posterior is also a Gaussian,
\begin{eqnarray}
	\mathcal{P}(s|d)&=&\mathcal{G}(s-m,D)\mbox{ with}\\
	D &=& \left(S^{-1}+R^\dagger N^{-1} R \right)^{-1},\\
	m &=& D\,j, \mbox{ and}\\
	j &=& R^\dagger N^{-1}d.
\end{eqnarray}
The signal posterior mean $m=\langle s \rangle_{(s|d)}$ is obtained from the data by projecting it into the data space as $j = R^\dagger N^{-1}d$ after inverse noise weighting and then by operating on it with the so called information propagator $D$. 
The operation that turns  the data into the signal estimate $m = F_\mathrm{W} d$, 
\begin{equation}
	F_\mathrm{W}=\left(S^{-1}+R^\dagger N^{-1} R \right)^{-1} R^\dagger N^{-1}\label{eq:WF},
\end{equation}
is called the Wiener filter \citep{1949wiener}.
The posterior signal uncertainty $\langle (s-m)\,(s-m)^\dagger  \rangle_{(s|d)}$ is also given by $D$, which is also called Wiener variance.\\
Eq. \ref{eq:WF} illustrates the complexity operators can have. 
They are composed of a number of operators that act on different spaces that often need to be inverted as well. 
Since the number of pixels used to represent fields can be large, an explicit storage of such operators is prohibitive. 
They have to be represented by computer routines that perform their actions. 
This, however, makes it difficult to access the entries of an operator, for example to examine the uncertainty of the Wiener filter reconstruction, which is encoded in the entries of $D$. \\
The building blocks of $D$, the signal and noise covariances $S$ and $N$, might be diagonal in harmonic spaces. 
In contrast, the response is usually best described in position space. 
Thus, the representation of $D^{-1}=S^{-1}+R^\dagger N^{-1} R $ requires harmonic transforms, e.g. Fourier transforms, and the application of $D$ to $j$ requires the solving of the linear system $D^{-1}m=j$, which asks for many evaluations of such transforms. \\
In figure \ref{fig:minimal_wiener_filter_demo} a minimal Wiener filter implementation in \niftyt\ is shown.
Thereby, an exemplary geometry is set up in the beginning and in the end the Wiener filter is applied explicitly. 
In the middle part the definition and initialization of the data (\verb|d|), the signal- and noise-covariances (\verb|S| and \verb|N|), and the response operator (\verb|R|) are left out, since this step is highly problem specific. 
Please refer to section \ref{sec:wiener_filter} for an exhaustive discussion of a Wiener filter implementation. 
 
\begin{figure*}[!t]
    \begin{lstlisting}[language=iPython]
    import nifty as ift                             # Import NIFTy package 
    from ift.library import WienerFilterCurvature   # Import Wiener filter from nifty.library
                                                    # Setting up...
    s_space = RGSpace([512, 512])                   #   ...position space (RGSpace as an example)
    fft = FFTOperator(s_space)                      #   ...an operator for harmonic transforms
    h_space = fft.target[0]                         #   ...harmonic space
    p_space = PowerSpace(h_space)                   #   ...power space
    
    d, S, N, R = ...                                # Define specific components here
    
    j = R.adjoint_times(N.inverse_times(d))         # Compute the "information source"
    C = WienerFilterCurvature(S=S, N=N, R=R)        # Setting up (the inverse of) the propagator
    m = C.inverse_times(j)                          # Apply the Wiener filter
    \end{lstlisting}
\caption{Minimal example for a Wiener filter reconstruction in two dimensions, whereby the instantiation of problem specific components is omitted.} 
\label{fig:minimal_wiener_filter_demo}
\end{figure*}

\subsection{Interacting Information Field Theory}\label{sec:interaction}
 
If any of the simplifying assumptions of the Wiener filter problem is violated, e.g.\ if signal or noise are not Gaussian, the noise depends on the signal, or one of the operators $R$, $S$, and $N$ is not known a priori, the inference problem becomes much more difficult and the posterior mean is not obtained by a linear transformation of the data. \\
However, even then an information Hamiltonian 
$\mathcal{H}(d,s) = -\ln \mathcal{P}(d,s)$ can be defined, which encodes all the information on the field. 
One might want to minimize this Hamiltonian with respect to the field $s$ to obtain the maximum a posteriori estimate for it. 
Or one uses this Hamiltonian within a variational Bayes scheme, in which an approximate posterior  $\widetilde{\mathcal{P}}(s|d)$ (usually a Gaussian) is matched to the full posterior via their Kullback-Leibler distance 
\begin{eqnarray}
	\mathrm{KL}(\widetilde{\mathcal{P}},\mathcal{P})
	&=&\int \mathcal{D}s\,\widetilde{\mathcal{P}}(s|d)\,
	\ln \left(\frac{\widetilde{\mathcal{P}}(s|d)}{{\mathcal{P}}(s|d)}\right)\nonumber\\
	&=&
	\left\langle
	\mathcal{H}(s|d)-\widetilde{\mathcal{H}}(s|d)
	\right\rangle_{\widetilde{\mathcal{P}}(s|d)}.
\end{eqnarray}
Therefore, dedicated numerical infrastructure is required to minimize such functionals efficiently.

\subsection{Manifold Independence \& Discretized Continuum}\label{sec:manifold_independence}
It is remarkable that the abstract algorithm for the inference of $s$ (e.g.\ equation \ref{eq:WF_condensed}) is independent of the choice of $\Omega$. 
In a concrete implementation, however, a discretization of $\Omega$ must be provided, since classical computers perform only discrete arithmetic. 
Still, because of this independence of $\Omega$, the implementation should be modular with respect to it. 
\niftyo\  was specifically designed for separating the inference algorithm from the problem's manifold. 
For this the concept of space- and field-objects was introduced:
	spaces describe the geometrical properties of physical space, 
	whereas fields are defined on spaces and carry actual data.
As a consequence, algorithms implemented within the \nifty\ framework are by design manifold and resolution independent. 
Particularly, \nifty\ takes care of correctly including \emph{volume factors} ($V$) whenever scalar products or, generally speaking, index contractions between tensors are performed. 
Volume factors are a direct consequence of the discretization of $\Omega$. 
Giving an example, the scalar product of two signal fields $s$ and $u$ on $\Omega$ behaves like 
\begin{equation}
s^\dagger u = \int_\Omega \mathrm{d}x \; s^*(x) u(x) \approx \sum_{q=1}^Q V_q^{\phantom{*}} s_q^* u_q^{\phantom{*}}
.
\end{equation}
Compare \citet[chapter 2.2]{2013A&A...554A..26S} for more details. \\

For completeness, two further key concepts of \nifty\ are discussed.

\subsection{Data Representation}\label{sec:data_representation}
For many signal reconstruction problems one can assume that the signal $s$ is statistically homogeneous and isotropic. 
This means that the statistical properties of $s(x)$ are independent of \emph{position} in $\Omega$ for the former, 
	and from \emph{direction} for the latter. 
Such fields have the remarkable property that their correlation structure can be described by a one-dimensional power spectrum which is diagonal in a harmonic basis of $\Omega$.
Hence, it is very beneficial to perform covariance-related calculations in the harmonic basis, 
since the involved Fourier transforms scale\footnote{Note that for regular grids \emph{fast} Fourier transforms exist that scale with the aforementioned $\mathcal{O}(N_{\mathrm{pix}} \log (N_{\mathrm{pix}}))$. 
On curved manifolds like the sphere and for non-regular spaced discretizations the costs may be higher.} like $\mathcal{O}(N_{\mathrm{pix}} \log (N_{\mathrm{pix}}))$
and therefore are cheap compared to full rank matrix multiplications scaling like $\mathcal{O}(N_{\mathrm{pix}}^2)$, where $N_{\mathrm{pix}}$ is the cardinality of $\Omega$'s pixelization. 
\nifty\ has representations for its supported manifolds in signal as well as harmonic bases, and 
transforming fields from one representation to another is done via efficient external libraries \citep{FFTW05, gorski2005healpix}.

\subsection{Implicit vs. Explicit Operators}\label{sec:implicit_explicit_operators}
In the previous section we have seen that in many cases one can choose a basis for which the covariance operators are diagonal. 
However, in general an arbitrary measurement device's response is not a square operator and therefore is not diagonalizable. 
Expressing the operator explicitly as a matrix of shape $(N_{\mathrm{pix}}, N_{\mathrm{data}})$ leads to problems if $N_{\mathrm{pix}}$ is large:
	first, matrix multiplications will slow down the reconstruction algorithm as they scale like $\mathcal{O}(N_{\mathrm{pix}}^2)$ and 
	second, the operator may simply not fit into main memory. 
Because of this, operators in \nifty\ are almost always stored as implicit objects. 
This means that instead of storing every entry of an operator's matrix representation explicitly, only its \emph{action} is implemented. 
However, expressing operators implicitly has the drawback that properties like the operator's diagonal, trace or determinant can -- within reasonable computing time -- only be determined approximately by probing \citep{Hutchinson1989Probing} that might be combined with an inference step \citep{2012PhRvE..85b1134S, 2015PhRvE..92a3302D}. 
Also, the inverse of an implicitly defined operator must be computed using numerical inversion algorithms like the conjugate gradient method \citep{Hestenes52methodsof}, whenever it is applied to a vector. 
To overcome those inconveniences, \nifty\ has the built-in functionality to compute those properties and actions transparently for the user.

\subsection{Reference Projects}
The \nifty\ framework provides rich structure in terms of a class hierarchy and built-in functionality,  while at the same time custom objects -- first and foremost operators -- can be implemented easily. 
Therefore, it is well suited for rapidly prototyping new inference algorithms, as well as for building full-grown signal reconstruction applications. 
Because of this, since its release in 2013, \niftyo\ has been used in various published scientific codes and papers. 
Some noteworthy examples are: 
\begin{itemize}
    \item Estimating galactic and extragalactic Faraday rotation \\ \citep{2012A&A...542A..93O-short, 2015A&A...575A.118O, 2016A&A...591A..13V}
	\item D$^3$PO for photo count imaging \citep{2015A&A...574A..74S}
    \item Signal inference with unknown response \citep{2015PhRvE..91a3311D}
    \item Cosmic expansion history reconstruction from SNe Ia data \\ \citep{Porqueres:2016kfv}
    \item Dynamic system classifier \citep{2016PhRvE..94a2132P}
    \item PySESA for geospatial data analysis \citep{2016CG.....86...92B}
	\item RESOLVE for radio interferometry \citep{2016A&A...586A..76J}
    \item Tomography of the Galactic free electron density \\ \citep{2016A&A...590A..59G}
    \item Bayesian weak lensing tomography \citep{2017arXiv170101886B}
    \item Noisy independent component analysis of auto-correlated components \citep{2017arXiv170502344K}
\end{itemize}

\section{Limitations of \niftyo}
Despite all its strengths, \niftyo\ has some considerable limitations which were the cause for developing \niftyt.
Those limitations are discussed in the following. 

\subsection{Combined Manifolds \& Field Types}\label{sec:combined_manifolds}
As discussed in section \ref{sec:manifold_independence}, \nifty\ separates the domain on which a field lives from the field itself.
The code architecture was designed for natively supporting only scalar fields on domains consisting of only one space.
However, there are many physical applications where this is too limiting.\\
For example, the D$^3$PO code \citep{2015A&A...574A..74S} performs a reconstruction of the diffuse and point-like $\gamma$-ray sky emissivity for a given set of photon counts.
This can be done for photon counts at arbitrary energies permitting the construction of sky images in distinct energy bands. 
However, the individual energy bands are reconstructed separately so far; i.e.\ for a fixed pixel and fixed energy band the relevant information from nearby energy bands is not exploited. 
The next logical step is to reconstruct and decompose the true sky emissivity for all energy bands jointly within one single reconstruction. 
In this scenario information about the energy correlation function should be used for linking and letting the information cross-talk between the energy bands. 
This involves fields living on a domain which is the combination, viz.\ the Cartesian product, of the geometrical space (the celestial sphere) and the energy dimension (a one-dimensional Cartesian space). \\
Another example is the reconstruction of the Galactic free electron density by \citet{2016A&A...590A..59G}. 
The algorithm was initially built for reconstructing scalar quantities.
But in principle the algorithm can be used for reconstructing vector fields, e.g.\ magnetic fields, or even tensor fields as well. 
This makes it necessary to be able to specify fields of arbitrary type; e.g. scalar, vector or tensor type. \\
Independently of its technical challenges, dealing with power spectra on domains being the Euclidean product of single spaces is a conceptually non-trivial task. 
Analyzing power spectra, as well as drawing random samples from given spectra for such domains, requires high diligence, as the statistical normalization and the field's Hermitianity are broken when using naive standard approaches. 
In section \ref{sec:domain_objects}, and especially in \ref{sec:power_analyze} and \ref{sec:power_synthesize} it is discussed how \niftyt\ tackles those challenges. 

\subsection{Scalability \& Parallelizability}\label{sec:scalability}
Signal reconstruction problems easily reach sizes where one mainly runs into two problems.
First, given a reconstruction algorithm, the individual steps of this algorithm scale at best linearly ($\mathcal{O}(N_{\mathrm{data}})$) with the size $N_{\mathrm{data}}$ of the data set.
For example, adding $1024~\mathrm{MiB}$ of array data takes twice as long as adding $512~\mathrm{MiB}$\footnote{For very small data arrays, a doubled array size can result in less than the doubled time because of constant-time overheads.}.
Fourier transforms scale with $\mathcal{O}(N_{\mathrm{data}}\mathrm{log} (N_{\mathrm{data}}))$ at best and explicit matrix-vector multiplications even only with $\mathcal{O}(N_{\mathrm{data}}^2)$.
Hence, with large data sets one easily reaches unacceptable computational run-times.\\
Second, for large problems the quantities of interest may not fit into a single computer's memory anymore. 
For example, for the reconstruction of the Milky Way's dust density in terms of a cubic volume, a side resolution of $2048$ is a reasonable degree of refinement. 
This cube contains $2048^3 \approx 8.6 \cdot 10^9$ voxels. 
Hence, a scalar field on this cube containing 64-bit floating point numbers consumes $64~\mathrm{GiB}$, and in practice one needs to handle several field instances at the same time. 
Thus, even if runtime was not a problem, the reconstruction for such a field simply does not fit onto an affordable shared memory machine.\\
Hence, \niftyo, which does not provide parallelization to the core, reaches its limits for large reconstruction problems. 
In section \ref{sec:parallelization} it is described how \niftyt\ makes use of the software packages \textsc{D2O} \citep{Steininger2016} to achieve scalability and parallelizability, 
and \emph{keepers}\footnote{\url{https://gitlab.mpcdf.mpg.de/ift/keepers}} for cluster compatibility. 

\subsection{Refactoring the Code Structure}
In addition to the issues addressed in the previous sections \ref{sec:combined_manifolds} and \ref{sec:scalability}, a whole set of further modifications was due, to ensure that the \nifty\ framework and the codes being built with it remain robust and modular in the future.  

\subsubsection{Energy Functionals} 
In almost all realistic applications an energy functional, for example the information Hamiltonian,
\begin{equation}
 \mathcal{H}(d,s) = - \ln \mathcal{P}(d, s)
\end{equation}
must be minimized to reconstruct a sought-after signal. 
See \citet{2015A&A...575A.118O}, \citet{2015A&A...574A..74S}, or \citet{2016A&A...590A..59G} as examples.
When minimizing the information Hamiltonian in the context of IFT, one has the advantage that its analytic form is known.
Hence, in contrast to most common minimization settings, the gradient, curvature and any other derivative of the energy functional are directly accessible and can be used by appropriate minimizers\footnote{The gradient can be used directly in gradient-based methods like \emph{steepest-descent} or \emph{LBFGS} \citep{Liu1989,Byrd:1995:LMA:210879.210980}; additionally, the true curvature can be used in Newton minimization schemes.}.
Because of this structure, evaluating the energy functional at a certain location naturally means to consecutively calculate the functional's derivatives with decreasing degree and reusing the previous information.\\
As an example, for a simple Wiener filter \citep{1949wiener} this looks as follows. 
Given a linear measurement of a signal $s \hookleftarrow \mathcal{G}(s, S)$ with an instrument having a response $R$ and additive noise $n \hookleftarrow \mathcal{G}(n, N)$, the data equation reads
\begin{equation}
d = R(s) + n
\end{equation}
(cf.\ equation \ref{eq:data-equation}).
The maximum-a-posteriori solution given by a Wiener filter is represented by the minimum of the following information Hamiltonian
\begin{equation}
\mathcal{H}(d, s) = \frac{1}{2}s^\dagger D^{-1} s - j^\dagger s  + \mathrm{const}
,
\end{equation}
where
\begin{equation}
D = (S^{-1} + R^\dagger N^{-1} R)^{-1} \qquad \mathrm{and} \qquad j = R^\dagger N^{-1} d 
.
\end{equation}
When evaluating $\mathcal{H}(x)$, one computes the components of its derivatives along the way, too:
\begin{equation}
\partial_{s_x} \partial_{s_y} \mathcal{H}(d, s) = (D^{-1})_{xy} \qquad \mathrm{and} \qquad \partial_{s_x} \mathcal{H} = D^{-1}s - j
\end{equation}
Hence, for computation efficiency it is crucial to reuse those very building blocks to avoid duplicate calculations. 
A minimizer, for example, will start with evaluating the functional's value at the given current position in parameter space.
\emph{Afterwards}, it will need the gradient and (depending on the algorithm) maybe the curvature.  
When storing the right portions of information, those queries can be answered immediately, speeding up the minimization process as a whole. \\
So far \nifty\ was missing structure which supported the efficient implementation of information Hamiltonians. 
This changed with \niftyt, cf.\ section \ref{sec:energy_object_and_minimization}.

\subsubsection{Modularity of Minimizers and Probers}
Real-life information Hamiltonians are hard to minimize since they may be ill-conditioned, multi-modal, non-convex and exhibit plateaus. 
This makes it all the more important to freely choose a minimization algorithm and maybe even sequentially combine several of them. 
Section \ref{sec:energy_object_and_minimization} shows how generality and modularity have been improved for \niftyt.\\
As described in section \ref{sec:implicit_explicit_operators}, probing is necessary to infer quantities like the trace or diagonal of an arbitrary implicitly represented operator. 
Since probing an operator can become the most costly part of an inference algorithm, it is crucial that this task is carried out wisely.
In section \ref{sec:implicit_explicit_operators} an overview is given how \niftyt\ supports the user in this respect. 

\subsubsection{Meta-Information on Power Spectra}
Power spectra are crucial quantities in most inferences that are built with \nifty, as they encode the statistical information which is used to overcome the limited and noisy nature of measurement data.  
However, in \niftyo\ power spectra are simply stored as un-augmented arrays, putting the burden on the users to manage all the meta-information like band structures, links to the harmonic partner space, or binning information.
In section \ref{sec:power_space} it is described how \niftyt\ uses the concept of PowerSpaces to confine all of this interrelated information in one place. 

\section{The Structure of \niftyt}
In this section we discuss how \niftyt\ meets the requirements stated in section \ref{sec:problem_description}.
\subsection{Domain Objects and Fields}\label{sec:domain_objects}
Fields are the main data carriers in \nifty. 
Beside its data array, a field instance also stores information about its data's domain.
A domain object can either be a \verb|Space| or \verb|FieldType| instance and it contains the information and functionality a field needs to know about the data it holds. 
A space in \niftyt\ defines a spatial geometry, or to be precise, it represents the finite resolution discretization of a certain continuous and compact manifold. 
Additionally, \niftyt\ also provides a field-type class,
which is the base class for all bundles that may be put on top of the manifold: vectors, tensors, or just lists of numbers\footnote{The \texttt{FieldArray} class represents a collection of numbers with a certain shape, but without any further structure.}.
This illustrates how the concept of spaces from \niftyo\ has been extended to a new base class called \verb|DomainObject|. 
The class tree is shown in figure \ref{fig:DomainObjects}. \\
A crucial feature of \niftyt\ is that the domain of a field is a \emph{tuple} of domain objects instead of a single space, i.e.\ in general the domain is the Cartesian product of spaces and field-types.
Actually, the \verb|domain| attribute of a field can contain an arbitrary amount of spaces and field types, including zero. 
This makes it a very general concept which naturally covers various use cases.\\
The conventional use case is a field which is defined over one single space, for example a two-dimensional regular grid space.

\begin{lstlisting}[language=iPython]
	|\iin| f = ift.Field(
               domain=ift.RGSpace(shape=(4, 4)),
    	       val=1))
    |\iin| f.domain 
    |\out| (RGSpace(shape=(4, 4), 
    		    	zerocenter=(False, False), 
    			    distances=(0.25, 0.25), 
        			harmonic=False),)    					 
	|\iin| f.val
	|\out| <distributed_data_object>
	       array([[1, 1, 1, 1],
				  [1, 1, 1, 1],
				  [1, 1, 1, 1],
				  [1, 1, 1, 1]])
\end{lstlisting}
Note that, as described above, for consistency with multi-domain fields \verb|f.domain| returns a \emph{tuple} of space(s) even though \verb|f| was defined on a single space.\\
The second example shows a field that is defined over the Cartesian product of a HEALPix sphere \citep{gorski2005healpix} and a one-dimensional regular grid space.

\begin{lstlisting}[language=iPython]
	|\iin| f = ift.Field(
            domain=(ift.HPSpace(nside=64),
                    ift.RGSpace(shape=(128,))))
	|\iin| f.shape
	|\out| (49152, 128)
	
	|\iin| f.dof
	|\out| 6291456
\end{lstlisting}
This can be used for a combined analysis of 128 energy bands for each pixel on a sphere.
Note that the shape of \verb|f|'s data array is composed of the shape of a single HEALPix sphere (49152) and the regular grid space. 
Hence, the field has 6291456 degrees of freedom in total.\\
The way to store data\footnote{Here, \emph{data} stands for the manifold-free result of a measurement in the sense of equation \ref{eq:measurement_equation}: $d = f(s, n)$.} in \niftyt\ is to define a field which lives on \emph{no} geometric space at all. 

\begin{lstlisting}[language=iPython]
	|\iin| f = ift.Field(
                 domain=ift.FieldArray((128,)))
	
	|\iin| f.shape
	|\out| (128,)
\end{lstlisting}
The domain concept generalizes so far that one can even define a field whose domain is empty:  

\begin{lstlisting}[language=iPython]
	|\iin| f = ift.Field(domain=(), val=123.)

	|\iin| f.shape
	|\out| ()

	|\iin| f.val
	|\out| <distributed_data_object>
		   array(123.)
\end{lstlisting}
This represents a single scalar, without any further information; neither geometrical nor structural.\\

\subsubsection{Separation of Responsibilities}
In contrast to \niftyo, domain objects in \niftyt\ only contain information and functionality regarding the geometry they define,
e.g., grid distances and pixel volumes. 
Actually, in \niftyo\ the field class was conceptually agnostic regarding the data object it possessed, as all numerical work was done by the space instances. 
In \niftyt\ fields are defined on a tuple of domain-objects and therefore numerical operations affect the structure as a whole. 
Hence, the responsibility for numerical operations had to be elevated from the space objects.
To be precise, this means that the responsibility for, e.g., arithmetics, power spectrum analysis, and random sample generation was moved into the field class. 
Functionality for, e.g., harmonic transforms, smoothing, and plotting was realized in terms of new dedicated operators. 
This change was an important step to better obey the \emph{single responsibility principle} \citep{martin2003agile}, which is why the code is now much more modular, stable and extendable.\\

The \nifty\ standard-library contains the following spaces. 

\subsubsection{RGSpace}

The \verb|RGSpace| ({\bf R}egular {\bf G}rid {\bf Space}) represents an $n$-dimensional toroidal manifold, i.e.\ the Cartesian product of $n$ circles ($S^1 \times ... \times S^1$).
Since the grid is regular, each pixel has the same volume weight, which in turn is given by the pixels' edge lengths, i.e.\ the \verb|distances| attribute. 
The harmonic partner space of an RGSpace is again an RGSpace with identical shape but different grid lengths. 
Actually, the harmonic space's grid is the reciprocal lattice of the signal space's grid.
Hence, the pixels' edge lengths in the harmonic grid are given by the inverse of the longest distances in the signal space grid\footnote{Since the \texttt{RGSpace} represents an n-dimensional torus, the longest distances are given by the circumferences of the individual circles the torus is made of.}.

\begin{lstlisting}[language=iPython]
	|\iin| signal_space = \
            ift.RGSpace(shape=(4, 5), 
                        distances=(0.8, 0.4))  
	|\iin| fft = ift.FFTOperator(signal_space)                       
	|\iin| f = ift.Field(signal_space, 
					      val=np.arange(20.))           
	|\iin| g = fft(f)                                            
	|\iin| g.domain                            
	|\out| (RGSpace(shape=(4, 5), 
                     zerocenter=(False, False), 
                     distances=(0.3125, 0.5),  
 					harmonic=True),)          
	|\iin| f.norm()                                              
	|\out| 28.114053425288926
	
	|\iin| g.norm()                                              
	|\out| 28.114053425288926
\end{lstlisting}
Note that the scalar product and therefore the norm remains invariant, irrespective of whether the field was represented in the signal or the harmonic basis.\\
Additionally, the \verb|RGSpace| class has the degree of freedom, whether the data should be stored \emph{zerocentered} or not. 
If the \verb|zerocenter| attribute is set to true for a certain axis, the origin is put at the center of the data array. 
This means that in case of a one-dimensional zero-centered harmonic space with $6$ pixels the values for $k$ are placed like:
\begin{equation}
\{k_{-3}, k_{-2}, k_{-1}, k_{0}, k_{1}, k_{2}\} \quad \mathrm{for} \quad k_i \quad \mathrm{with} \quad i \in [-3, ..., 2]
\end{equation}
In contrast, non-zero-centered ordering yields
\begin{equation}
\{k_{0}, k_{1}, k_{2}, k_{-3}, k_{-2}, k_{-1}\} \quad \mathrm{for} \quad k_i \quad \mathrm{with} \quad i \in [-3, ..., 2]
\end{equation}
Albeit being less convenient when looking at a spectrum by eye, fft libraries like \emph{FFTW} \citep{FFTW05} usually follow the non-zero-centered convention. 
\niftyt\ can handle both conventions and performs the mediation with respect to those external libraries.

\subsubsection{HPSpace}

The \verb|HPSpace| ({\bf H}EAL{\bf P}ix {\bf Space} \citep{gorski2005healpix}) represents a unit 2-sphere. 
The HEALPix pixelation subdivides the surface of a sphere into pixels of equal area. 
Those pixels reside on iso-latitude circles, with equal spacing on each circle. 
In \nifty\ the represented sphere has a fixed radius of $1$. 
Hence, the only remaining parameter for \verb|HPSpace| is the HEALPix resolution parameter \verb|nside|.
The harmonic partner space for \verb|HPSpace| is the \verb|LMSpace|.

\subsubsection{GLSpace}

Similar to the \verb|HPSpace|, the GLSpace ({\bf G}auss-{\bf L}egendre {\bf Space}) is a discretization of the unit 2-sphere.
It consists of $n_\text{lat}$ iso-latitude rings containing $n_\text{lon}$ pixels each. 
The pixels are equidistant in azimuth on every ring, and the ring latitudes coincide with the roots of the Legendre polynomial of degree $n_\text{lat}$.
Within each ring the pixel weights are identical.
However, the individual weights for the different rings are chosen such that a numerical integration is a Gauss quadrature that is exact for all spherical harmonics $Y_{lm}$ with $l<n_\text{lat}$ and $|m|\le (n_\text{lon}-1)/2$.
The harmonic partner of a \verb|GLSpace| is the \verb|LMSpace|. 

\subsubsection{LMSpace}

The \verb|LMSpace| is the harmonic partner domain for a pixelization of the unit 2-sphere. 
Its name is derived from the typically employed indices of the spherical harmonics $Y_{lm}$.
A \verb|LMSpace| instance holds a value for every $Y_{lm}$ with $-l\leq m\leq l$ and $0\leq l\leq l_\text{max}$ where $l_\text{max}$ is the cutoff frequency. 
It can be a harmonic partner to either the HPSpace or the GLSpace. 
There is a unique projection from LMSpace to HPSpace and to GLSpace, respectively.
However, this projection is not necessarily invertible; see section \ref{sec:fft} for details.
The volume weight is $1$ for each of the pixels in  \verb|LMSpace|.

\subsubsection{PowerSpace}\label{sec:power_space}

The \verb|PowerSpace| is used to condense the $k$-modes of a harmonic space into a one-dimensional space, the discrete isotropic power spectrum.
This is a natural step when analyzing statistically homogeneous and isotropic fields,
	and the \verb|PowerSpace| class provides a rich functionality for this kind of analysis (cf.\ section \ref{sec:power_analyze}).
The \verb|PowerSpace| condenses the information of a harmonic space of arbitrary dimension into a one-dimensional space by defining bins into which the different $k$-modes are aggregated.
By this, it enables data scientists to write algorithms that are independent of the underlying space.
The \verb|PowerSpace| keeps information about the relation to its harmonic partner space for both numerical efficiency and enabling field synthetization (cf.\ section \ref{sec:power_synthesize}).\\
Note that depending on the application there is an ambiguity to what the correct volume weighting for this space should be, cf.\ section \ref{sec:manifold_independence}.
The first weighting is induced by the harmonic partner space and uses the multiplicity $\rho$ of the modes in each bin as weights.
This weighting is highly dependent on the structure of the harmonic partner space.
The second weighting is induced by regarding the power space as a one-dimensional space on a (potentially) irregular grid, which induces a volume factor equal to the bin size.
The third weighting is obtained analogously to the second but differs by using a logarithmic axis $\mathrm{log}(k)$.
It is used when calculating smoothness on a log-log scale, which is required when a prior that favors power law spectra is wanted (cf.\ \citealt{ensslin2011reconstruction}).
Finally, there is a trivial fourth weighting where the volume factors are constantly set to one. 
This is, for example, needed in the critical filter (cf.\ \citealt{ensslin2011reconstruction}).
	
\subsubsection{Power Spectrum Analysis}\label{sec:power_analyze}
In this section we describe how the one-dimensional power spectrum for a given field is computed.
As mentioned in section \ref{sec:data_representation}, this one-dimensional quantity fully describes the correlation structure of statistically homogeneous and isotropic fields. 
The reason for this is that for such fields the statistical properties in the harmonic space solely depend on the distance to the origin, but not the direction. \\
The following example is in order:
Let \verb|s| be the discretization of a two-dimensional Cartesian space and \verb|f| a field on \verb|s|:
\begin{lstlisting}[language=iPython] 
	|\iin| s = ift.RGSpace(shape=(4,4),
					        zerocenter=True)
	|\iin| s 
	|\out| RGSpace(shape=(4, 4), 
                    zerocenter=(True, True), 
				    distances=(0.25, 0.25), 
				    harmonic=False)
	|\iin| f = ift.Field(domain=s, val=1.)
\end{lstlisting}
\verb|s| has a total volume of $1$ by default, which is the reason why the edge length of \verb|s|'s pixels is equal to $0.25$.  
To compute its power spectrum, \verb|f| first must be transformed to the harmonic basis.
\begin{lstlisting}[language=iPython] 
	|\iin| fft = ift.FFTOperator(s)
	|\iin| f_h = fft(f)
	|\iin| s_h = f_h.domain[0]
    |\iin| s_h
	|\out| RGSpace(shape=(4, 4), 
	                zerocenter=(True, True),  
	                distances=(1.0, 1.0), 
	                harmonic=True)
\end{lstlisting}
Now, the power spectrum can be computed by:
\begin{lstlisting}[language=iPython] 
	|\iin| f_p = f_h.power_analyze()
\end{lstlisting}
Note that the domain of \verb|f| is a \verb|PowerSpace| instance, viz.\ a one-dimensional irregularly gridded space with $6$ pixels:
\begin{lstlisting}[language=iPython] 
	|\iin| s_p = f_p.domain[0]
	|\iin| s_p
	|\out| PowerSpace(
        harmonic_partner=
			RGSpace(shape=(4, 4), 
			        zerocenter=(True, True), 
					distances=(1.0, 1.0), 
					harmonic=True), 
		distribution_strategy='not', 
		logarithmic=False, 
		nbin=None, 
        binbounds=None)
	|\iin| f_p.shape
	|\out| (6,)
\end{lstlisting}
Let's discuss what happened during the \verb|f_h.power_analyze()| call. 
Recall that \verb|f_h| is defined on a $4 \times 4$ grid. 
First of all, \verb|s_h| is asked for the distances with respect to the center:
\begin{lstlisting}[language=iPython] 
	|\iin| s_h.get_distance_array('not')
	|\out| 
	<distributed_data_object>
	array([[ 2.82,  2.23,  2. ,  2.23],
    	   [ 2.23,  1.41,  1. ,  1.41],
		   [ 2.  ,  1.  ,  0. ,  1.  ],
		   [ 2.23,  1.41,  1. ,  1.41]])

\end{lstlisting}
There are several pixels which have the same distance to the center. 
Those pixels together constitute a bin; hence, there are six bins in total. 
A pixel's bin-affiliation is stored in the so-called \verb|pindex| array:
\begin{lstlisting}[language=iPython] 
	|\iin| s_p.pindex
	|\out| <distributed_data_object>
            array([[5, 4, 3, 4],
        	       [4, 2, 1, 2],
        	       [3, 1, 0, 1],
        	       [4, 2, 1, 2]])
\end{lstlisting}
The multiplicity of each bin is given by \verb|rho|; its distance to the grid center by \verb|kindex|:
\begin{lstlisting}[language=iPython] 
	|\iin| s_p.rho 
	|\out| array([1, 4, 4, 2, 4, 1])
    |\iin| s_p.kindex
    |\out| 
    array([0., 1., 1.41, 2., 2.24, 2.83])
    
\end{lstlisting}
Concretely, calculating the power spectrum of \verb|f_h| now means accumulating the absolute squared of \verb|f_h|'s pixel values in their respective bins and divide by 
\verb|rho| to form the average. 
Assuming the artificial case that \verb|f_h| has the following values
\begin{lstlisting}[language=iPython] 
	|\iin| f_h.val
	|\out| 
	<distributed_data_object>
	array([[-1.+3.j, 4.+5.j, 2.+4.j, 0.+2.j],
           [-3.+6.j,-3.+1.j,-4.+2.j, 3.+3.j],
		   [ 1.+4.j, 2.+7.j,-4.+1.j, 4.+5.j],
		   [ 3.+5.j,-2.+0.j,-3.+6.j,-3.+1.j]])
\end{lstlisting}
this results in
\begin{lstlisting}[language=iPython] 
	|\iin| f_p = f_h.power_analyze()
	|\iin| f_p.val
	|\out|
	<distributed_data_object>
	array([ 17., 39.75, 10.5, 18.5, 31., 10.])
\end{lstlisting}
Note that the power spectra we have discussed so far are always real. 
However, if a field's domain is the Cartesian product of multiple spaces, it can be necessary to keep track of the real and imaginary part of \verb|f| in the position basis separately. 
In \niftyt , this can be controlled with the \verb|keep_phase_information| keyword in \verb|power_analyze|. 
If it is set to true, the resulting power spectrum will be complex. 
Its real (imaginary) part is constructed from the real (imaginary) part of \verb|f|. 
To avoid harmonic transforms whenever possible, the decomposition is done using Hermitian symmetry constraints, see section \ref{sec:hermitian_decomposition} for details. 

\subsubsection{Power Spectrum Synthesis}\label{sec:power_synthesize}
It is very important to be able to create random samples for a given power spectrum. 
For a field \verb|f_p| in \niftyt , this can be done using the method \verb|f_p.power_synthesize()|. 
This routine draws a Gaussian random field with zero mean and unit variance in the harmonic partner space of the power space.
Afterwards, the modes are weighted according to the power spectrum that is stored in \verb|f_p|. \\
Note that it is non-trivial to create the Gaussian random field sample in case of real-valued signal-space fields.
This is because the result of Fourier-transforming a real-valued signal space field (\verb|f|) has a point symmetry with respect to the origin in the harmonic basis.
This becomes particularly exigent if one synthesizes a random field from a multi-dimensional power spectrum which is associated with the Cartesian product of two or more spaces. 
In this case one needs a well-engineered approach for handling the aforementioned field's symmetry. 
This so-called \emph{Hermitian decomposition} is discussed in the next section. 

\subsubsection{Hermitian Decomposition}\label{sec:hermitian_decomposition}
Hermitian decomposition is needed for both, power spectrum analysis and synthesis. 
Hence, in the following its rationale is discussed.\\
The result of a harmonic transform of a real-valued field is in general complex-valued.\footnote{This is true if the customary transformation kernels are used;
	e.g.\ for flat manifolds this is $e^{-2 \pi i k x}$. 
	However, in principle also sine and cosine kernels are applicable, where the real/imaginary part of the field in position space gets mapped onto the real/imaginary part of the field in harmonic space, respectively.}
However, the degrees of freedom are the same for both representations, since -- as noticed in the previous section -- in the harmonic basis the field exhibits a point symmetry modulo complex conjugation. 
Therefore, it is called \emph{Hermitian} symmetry.  
Because of the complex conjugation, the imaginary part of the reflection's fixed points vanishes. 
For power spectrum analysis we need the functionality to decompose a field into its Hermitian and anti-Hermitian part.
In addition, for power-spectrum synthesis of real fields, one needs normally distributed Hermitian random samples. \\
In the following example the harmonic representation of a real-valued random field is shown. 
The fixed points of the Hermitian symmetry, where only real numbers should appear, are marked red. 
\begin{lstlisting}[language=iPython] 
    |\iin| s = ift.RGSpace((4, 4))
    |\iin| f = ift.Field(
                s, val=np.random.random((4, 4)))
    |\iin| f *= 10.
    |\iin| f.val
    |\out| <distributed_data_object>
            array([[0.7, 9.5, 9.7, 8.1],
                   [0.3, 1.0, 6.8, 4.4],
                   [1.2, 5.0, 0.3, 9.1],
                   [2.6, 6.6, 3.1, 5.2]])
    |\iin| fft = ift.FFTOperator(s)
    |\iin| fft(f).val
    <distributed_data_object>
    array(
    [[|\red{4.8+0.j}|  ,-0.8+0.3j,|\red{-1.3+0.j}|  ,-0.8-0.3j],
     [0.8+0.1j,-0.3-0.1j, 0.3-0.7j,-0.9+0.6j],
     [|\red{0.7+0.j}|  ,-0.2+0.j ,|\red{-1.1+0.j}|  ,-0.2-0.j ],
     [0.8-0.1j,-0.9-0.6j, 0.3+0.7j,-0.3+0.1j]])
\end{lstlisting}	
Please note that, for example, the entry at position $(2,4)$ with value ($-0.9 + 0.6\iu$) is the complex conjugated partner of position $(4,2)$ with value ($-0.9 - 0.6\iu$). \\
As mentioned in section \ref{sec:power_analyze}, for power-spectrum analysis it can be necessary to decompose a complex-valued position-space field into its real and imaginary part. 
However, for signal inference often the native basis is a field's harmonic space. 
In this case, one certainly could transform the field from its harmonic into its position space basis, split real and imaginary parts there, and, finally, transform back. 
Nevertheless, one ought to avoid harmonic transforms whenever possible, since they are at least either computationally expensive, or even inherently inexact, cf.\ section \ref{sec:fft}.
Luckily, as seen in the example above, the real part of a position-space field corresponds to the Hermitian symmetric part of the harmonic-space partner field. 
Since harmonic transforms are linear operations, the imaginary part corresponds to the anti-symmetric part of the field, analogously. 
Hence, splitting a position-space field into real and imaginary part is equivalent to splitting the harmonic-space field into its Hermitian and anti-Hermitian part. \\
For synthetization of fields from power spectra, as discussed in \ref{sec:power_synthesize}, it is necessary to create a normal random field that lives in harmonic space and has zero mean and unit variance. 
If the position space field shall be real, there are two approaches that correspond to those of the former paragraph. 
First, a real-valued sample can be drawn in position space which is then transformed into the harmonic basis. 
For the reasons set out above, this approach is sub-optimal. 
Second, a random sample can be created in harmonic space directly, from which the anti-Hermitian part is removed. 
By this, again, any harmonic transforms are avoided. 
However, particular diligence is needed in order to preserve the correct variances. \\
In the following it is discussed how a field can be decomposed into its (anti-)Hermitian parts. 
A straightforward approach is to use a reflection or flipping operator $\gamma$ which performs a point reflection with respect to a space's center.
Therefore for a given field $f$, the Hermitian and anti-Hermitian parts $f_h$ and $f_a$, respectively, are given by the half-sum and half-difference
\begin{eqnarray}
f_h &=& \frac{1}{2} \left(f + \gamma(f)^*\right)\\
f_a &=& \frac{1}{2} \left(f - \gamma(f)^*\right)
,
\end{eqnarray}
where
\begin{equation}
f = f_h + f_a
. 
\end{equation}
This works perfectly fine for individual spaces and also generalizes for the Cartesian products of spaces. 
Assuming a field $f$ whose domain is $(s_1, s_2)$, the individual flips must be applied consecutively before applying the complex conjugation. 
\begin{eqnarray}
f_h &=& \frac{1}{2} \left(f + \gamma_1(\gamma_2(f))^*\right)\\
f_a &=& \frac{1}{2} \left(f - \gamma_1(\gamma_2(f))^*\right)
\end{eqnarray}

\paragraph{Correcting the variance}\label{sec:correcting_the_variance}
When drawing samples for a given power spectrum it is necessary to create random samples with zero mean and unit variance. 
However, when forming the half-sums and -differences of a Gaussian random field with the mirrored version of itself, its variance must be corrected by a factor of $\sqrt{2}$\footnote{Recap that the sum of two Gaussian distributions is again a Gaussian distribution.}.
However, a priori the correction only applies to those pixels that got mapped to different pixels via the flipping. 
The pixels which the flip maps onto themselves, i.e.\ the fixed points of the flip, are averaged with themselves and therefore do not need a variance correction factor. 
But once the sample is complex valued, the involved complex conjugation will eliminate the fixed point's imaginary part in case of the Hermitian portion. 
To compensate this loss of power, in this case the fixed points after all need a correction factor of $\sqrt{2}$, too. 

\subsection{Linear Operators}
Up to now we have introduced fields as well as their domain-objects. 
In the following paragraph we will show how linear and implicitly defined operators act on fields and how they are implemented.\\
Every operator in \nifty\ is obliged to be linear and therefore inherits from the abstract base class \verb|LinearOperator|.
This base class provides the user with a blueprint for any elaborated operator. 
A generic operator $A$ is defined by its input and target domains, the boolean whether it is unitary or not, and its actual action. 
With these properties at hand, the application of $A$ on a given field $s$, 
	viz.\ $A\left(s\right)$, is split into a generic and a specific part. 
The generic part involves consistency checks, e.g.\ whether the domain of $s$ and $A$ match, while the specific part is the individual operator's action.
More specifically, the application of $A\left(s\right)$ is implemented as \verb|A(s)| or equivalently as \verb|A.times(s)|. 
Here, \verb|times| basically first calls the internal method \verb|_check_input_compatibility|, 
	which performs the domain compatibility check for $s$ and $A$.
This is followed by the implementation of the operator's action in \verb|_times|. \\
In case $s$ is defined over multiple domain-objects but $A$ is not, 
	and $A$ is supposed to act on a specific domain-object of $s$, 
	one pins the action of $A$ to a certain space of $s$'s domain tuple.
This is done by passing the information along with \verb|times|, e.g. \verb|A.times(s, spaces=0)|, or by using the \verb|default_spaces| keyword argument during the operator's initialization. 
The output of $A(s)$ may live on a different domain, depending on the target of $A$. \\
As one often needs more than the pure forward application of an operator on a field, i.e. \verb|.times|, \nifty\ allows the user to further implement \verb|_adjoint_times|, \verb|_inverse_times|, and \verb|_adjoint_inverse_times| if needed. 
In case $A$ is unitary, one only needs to implement \verb|_inverse_times| or \verb|_adjoint_times|, as \niftyt\ does a mapping of methods internally. 
\niftyt\ provides the user with a set of often reappearing operators. 
An overview can be found in table \ref{tab:operators} and figure \ref{fig:Operators}. \\

In the following, two especially mentionable operators will be described: the \verb|FFTOperator| and \verb|SmoothingOperator|.

\begin{table*}[!t]
	\caption{Overview of provided Operators, inherited from \texttt{LinearOperator}}
	\centering
	\begin{tabular*}{\textwidth}{@{\extracolsep{\fill}}|ll|}
		\hline
		\multicolumn{1}{|c}{\textsc{NIFTy} Operator} & \multicolumn{1}{c|}{description} \\
		\hline
		\hline
  		$\hookrightarrow$ \verb|ComposedOperator| & represents a container of \verb|LinearOperators| which are applied in sequence.\\
		$\hookrightarrow$ \verb|EndomorphicOperator| & represents all operators with equal domain and target.\\
		$\phantom{\hookrightarrow} \hookrightarrow$ \verb|DiagonalOperator| & represents a diagonal matrix in specified domain.\\
  		$\phantom{\hookrightarrow} \hookrightarrow$ \verb|LaplaceOperator| & represents the discrete second derivative for 1D spaces.\\
		$\phantom{\hookrightarrow} \hookrightarrow$ \verb|ProjectionOperator| & projects the input onto a \nifty\ field.\\
  		$\phantom{\hookrightarrow} \hookrightarrow$ \verb|SmoothingOperator| & convolves the input with a Gaussian kernel.\\
        $\phantom{\hookrightarrow} \hookrightarrow$ \verb|SmoothnessOperator| & measures the smoothness of the input by applying the LaplaceOperator twice.\\
		$\hookrightarrow$ \verb|FFTOperator| & implementation of harmonic transforms of fields between different domains.\\
		$\hookrightarrow$ \verb|ResponseOperator| & represents an exemplary response including convolutional smoothing and exposure.\\
		\hline
	\end{tabular*}
	\label{tab:operators}
\end{table*}

\subsubsection{FFTOperator}\label{sec:fft}

The purpose of this operator is the conversion of data from the spatial domain to the frequency domain and vice versa. 
For data living on equidistant Cartesian grids (i.e. RGSpaces) this is more or less straightforward to do:
every configuration in position or frequency domain has a unique corresponding configuration in the other domain.
The conversion between both sides is done efficiently -- and without information loss -- using an FFT algorithm. \\
For data living on the 2-sphere, things are more complicated: first of all, it is not possible to define pairs of equivalent spaces in the sense that data conversion is lossless in both directions for arbitrary maps. 
However, for any given LMSpace $L$ a GLSpace $G$ can be found, such that any data set living on $L$ can be converted to $G$ and back without losing information\footnote{Numerical noise caused by the finite precision of floating-point algebra can slightly break invertibility, though.}. 
In contrast, it is not possible to find an LMSpace that can hold the harmonic equivalent of all possible data sets living on a GLSpace. 
Hence, the set of \emph{band-limited} functions on the sphere, which can be transformed without loss in both directions, is only a tiny subset of all possible functions that can live on a GLSpace grid.
When solving problems on the sphere, it is therefore advantageous to formulate them primarily in harmonic space. \\
For the HEALPix discretization of the sphere, things are somewhat worse still: since no analytic quadrature rule for this set of pixels exists, any transform from LMSpace to HPSpace and back will only produce an approximation of the input data. 
In many situations this is acceptable given the other advantages of this pixelization, like equal-area pixels.
But this imperfection must be kept in mind when choosing pixelizations on which to solve a problem. \\
In \niftyo\ harmonic transformations are performed using a method of the particular field instance \verb|f|, viz.\ \verb|f.transform()|. 
This is a source of confusion and suggests general invertibility of the harmonic transform, since one could naively assume that \verb|f.transform().transform()| is equal to \verb|f| -- which is, incidentally, actually true for RGSpaces.  
In \niftyt\ the \verb|FFTOperator| solely implements the \verb|times| and \verb|adjoint_times| methods. 
The \verb|inverse_times| is available if and only if a certain transform path is explicitly marked to be unitary. 
This structure makes it very apparent when there is the risk of information loss. 

\subsubsection{Smoothing Operator} \label{sec:smoothing_operator}
This operator convolves a field with a Gaussian smoothing kernel.
The preferred approach is to apply the convolution in position space as a point-wise multiplication in harmonic space if possible.
The \verb|PowerSpace| is in general non-regularly gridded, which makes harmonic transformations hard, albeit not impossible.
But it is not periodic either, which is why convolutional smoothing cannot be applied to it. 
For this case, the \verb|SmoothingOperator| will perform its smoothing explicitly in position space. 

\subsubsection{Harmonic Smoothing Operator} \label{sec:harmonic_smoothing}
This operator exploits the fact that a convolution in the spatial domain is equivalent to multiplication in the frequency domain. 
Hence, the input field is first transformed using the \verb|FFTOperator|. 
The Gaussian kernel to be multiplied is generated with a standard deviation $\sigma$, given in the units of the domain's geometry. 
The result after multiplying the transformed input data is reverted to the position domain by calling \verb|FFTOperator|'s \verb|adjoint_times| method.
Thus, the caveats pertaining to transformations mentioned in the \ref{sec:fft} for \verb|FFTOperator| apply here as well.\\
Convolutional smoothing is available for all position spaces, i.e. \verb|RGSpace|, \verb|GLSpace| and \verb|HPSpace|. 
Internally, a method called \verb|get_fft_smoothing_kernel_function| is needed for the respective harmonic partner domain; it returns a function which is used to construct the kernel and is therefore defined on \verb|RGSpace| and \verb|LMSpace|. 
Note that the correct Gaussian kernel function for smoothing on the sphere is given by
\begin{equation}
K_{\text{sphere}}(l) = \exp(-\frac{1}{2} l(l + 1) \sigma^2)
\end{equation}
instead of the standard frequency response of a Gaussian filter
\begin{equation}
K(k) = \exp(-2 (\pi k \sigma)^2)
,
\end{equation}
since one has to take into account the sphere's curvature (cf.\ \citealt{challinor2000allsky}).

\subsubsection{Direct Smoothing Operator}
\begin{figure*}
\begin{subfigure}{.5\textwidth}
  \centering
  \includegraphics[width=.98\linewidth]{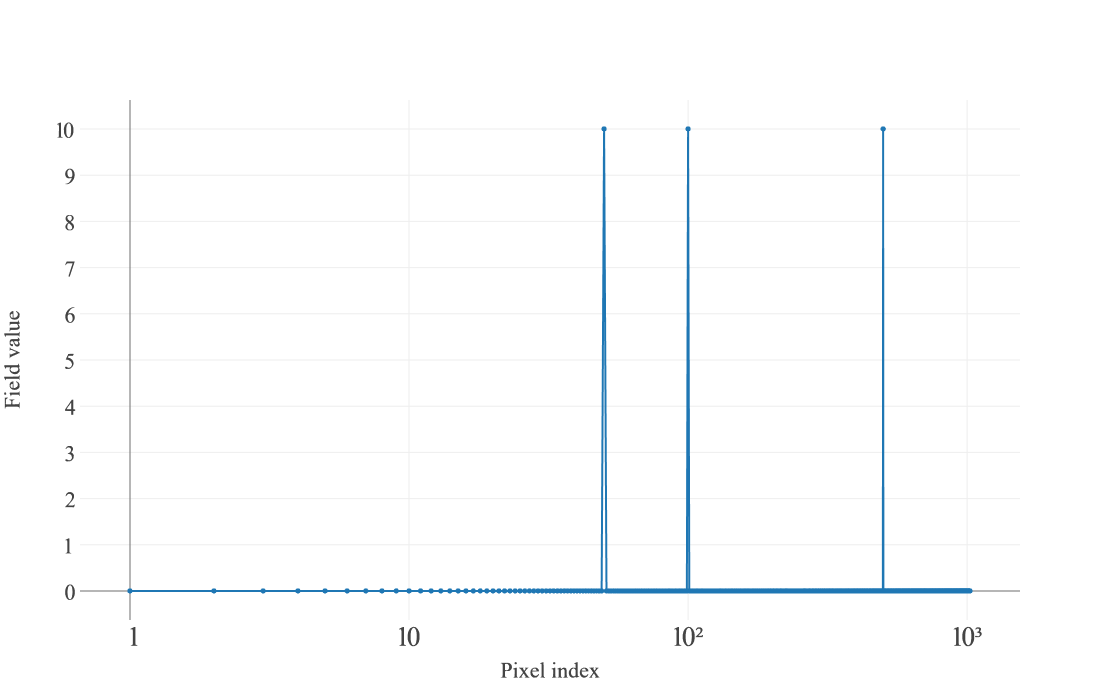}
  \caption{Unsmoothed field with three peaks.}
%  \label{fig:sfig1}
\end{subfigure}%
\begin{subfigure}{.5\textwidth}
  \centering
  \includegraphics[width=.98\linewidth]{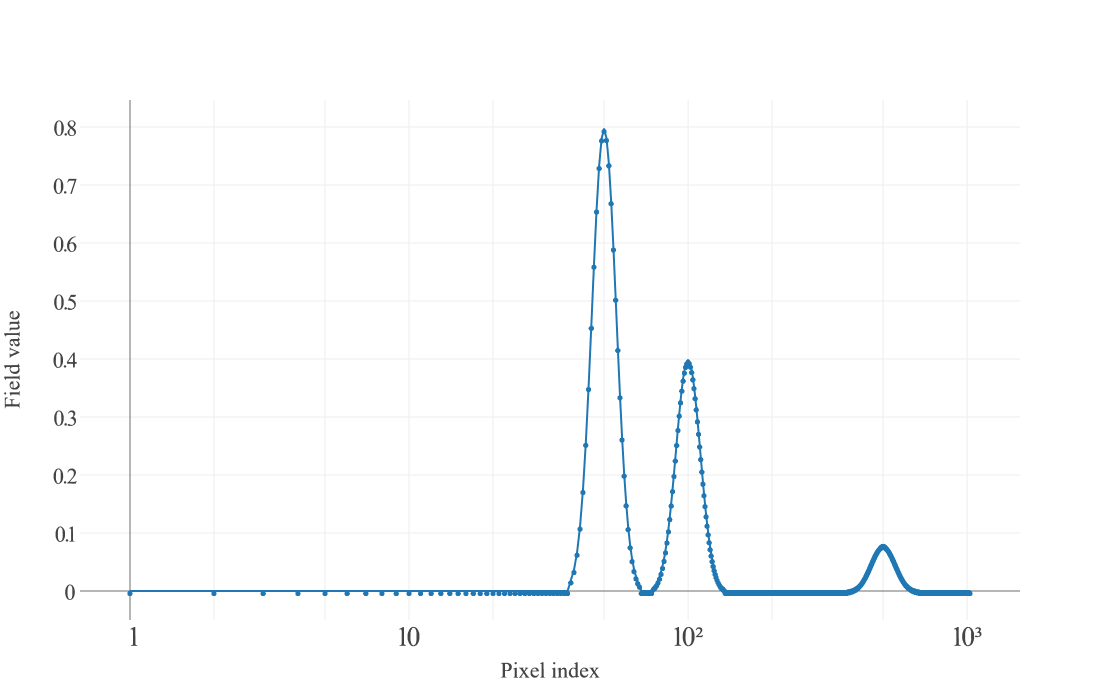}
  \caption{Smoothed field with different height for each peak.}
%  \label{fig:sfig2}
\end{subfigure}
\caption{Illustration of the behavior of smoothing an input field (a) on a non-regular grid or on a non-regular scale, respectively, which yields (b). Here, the field was smoothed on a logarithmic scale.}
\label{fig:power_smoothing}
\end{figure*}

The main motivation behind this operator is the aforementioned \verb|PowerSpace| from section \ref{sec:power_space}, which is based on a non-regular grid.
Since it is not periodic, this approach does not do a transformation into harmonic space but applies smoothing directly.
A point-by-point convolution is done on the space with a 1-dimensional Gaussian kernel with standard deviation $\sigma$. \\
In a normal case of convolution with a kernel of window length $w$ (assume as \textit{odd} for centered symmetric kernels) the result is shorter than the original data by $w-1$ points, losing $(w-1)/2$ points on each end of the data vector. 
Since this behavior is unsuitable as a solution for \nifty, the convolution at the beginning and end of the array, is carried out with partial Gaussian kernels.
Although this method is less accurate at the ends of the data, it does not sacrifice information resolution at each smoothing.
Also note that, since the \verb|SmoothingOperator| must conserve the sum of all field entries, on a non-regular grid the exact smoothing behavior depends on the location within the grid. 
The denser a certain pixel's neighborhood, the larger the number of weight receptors and therefore the lower the resulting Gaussian bell, when applying the \verb|SmoothingOperator| on single peaks. 
An illustration for this behavior is given in figure \ref{fig:power_smoothing}.
There, the result of smoothing a constant field with three dedicated peaks is shown.  
Since the smoothing happens on a logarithmic scale, the peak's weight is distributed to a different number of pixels for each peak. 
The code that was used to produce the plots in figure \ref{fig:power_smoothing} is given by
\begin{lstlisting}[language=iPython] 
    import nifty as ift
    power_space = ift.PowerSpace(
                    ift.RGSpace((2048), 
                                harmonic=True))
    f = ift.Field(power_space, val=1.)
    f.val[[50, 100, 500]] = 10
    sm = ift.SmoothingOperator(
                            power_space, 
                            sigma=0.1, 
                            log_distances=True)
    plotter = ift.plotting.PowerPlotter()
    plotter.figure.xaxis.label = 'Pixel index'
    plotter.figure.yaxis.label = 'Field value'
    plotter(f, path='unsmoothed.html')
    plotter(sm(f), path='smoothed.html')
\end{lstlisting}	

\subsection{Operator Inversion}
There are cases in which the inverse action of an operator is needed, but not analytically accessible and therefore not directly implementable. 
An example for this is the propagator operator in the Wiener filter (cf.\ equation \ref{eq:WF}), where the quantity $D^{-1}=\left(S^{-1}+R^\dagger N^{-1} R \right)$ must be inverted. 
For such cases \niftyt\ provides a robust implementation of the conjugate gradient method, following \citet{nocedal2006numerical}.\\
Additionally, \niftyt\ supplies the user with the special class \verb|InvertibleOperatorMixin|. 
In conjunction with multiple inheritance, this mixin-class (cf.\ \citealt[p. 599]{lutz2010programming}), can be used to equip a custom operator with missing \verb|*_times| methods. 
For example, if solely the \verb|inverse_times| of a custom operator is implemented, the mixin-class will provide the \verb|times| via conjugate gradient. 

\subsection{Probing}\label{sec:probing}
\niftyo\ already provided a class for probing the trace and diagonal of an implicitly defined operator $A$ by evaluating the following expressions:

\begin{align}
	\mathrm{tr}[A] &\approx \left< \xi^\dagger A \xi \right>_{\{\xi\}}
	= \sum_{pq} A_{pq} \left< \xi_p \xi_q \right>_{\{\xi\}} \rightarrow \sum_{p} A_{pp}
	, \\
	\label{eq:diag}
	\Big( \mathrm{diag}[A] \Big)_p &\approx \left( \left< \xi \ast A \xi \right>_{\{\xi\}} \right)_p
	= \sum_q A_{pq} \left< \xi_p \xi_q \right>_{\{\xi\}} \rightarrow A_{pp}
	,
\end{align}
where $\left<\,\cdot\,\right>_{\{\xi\}}$ is the sample average of fields $\xi$ which have the property $\left< \xi_p \xi_q \right>_{\{\xi\}} \rightarrow \delta_{pq}$ for $\#\{\xi\} \rightarrow \infty$, and $\ast$ denotes component-wise multiplication \citep[sec. 3.4]{2013A&A...554A..26S}. 
However, especially if the operator evaluation involves a conjugate gradient, the probing can account for the majority of an algorithm's computational costs, which is why several heuristics turned out to be useful in practice. 
Namely, in iterative schemes, it can be beneficial to reuse some or even all probes $\{\xi\}$ and/or utilize the former results as starting points for the conjugate gradient runs.  
To naturally allow for such heuristics, the prober class has been completely rewritten for \niftyt: now it exhibits special call-hooks for customization. 

\subsection{Energy Object \& Minimization}\label{sec:energy_object_and_minimization}
Inferring posterior parameters typically requires the minimization of some energy functional, such as the Information Hamiltonian or a Kullback-Leibler divergence.
The \verb|Energy| class provides the structure required for efficient implementation of such functionals.
Its interface provides the current value, its gradient and curvature at its position.
To avoid multiple evaluations of the same quantity, intermediate results can be memorized and reused later.
To ensure consistency when reusing intermediate results, an \verb|Energy| class instance in \niftyt\ is fixed to one certain position.
Setting a new position therefore creates a new instance of the class.
\verb|Energy| instances are passed to minimizers, which perform the actual numerical minimization.
The position of the passed instance is used as the starting point of the minimization.\\
\niftyt\ provides implementations for steepest descent, VL-BFGS \citep{NIPS2014_5333} and a relaxed Newton scheme.
Those methods differ in how the descent direction is determined.
Steepest descent just follows the downhill gradient.
VL-BFGS incorporates prior steps to estimate the curvature of the energy landscape, suggesting an improved direction.
The relaxed Newton minimizer makes use of the full local curvature, additionally providing an estimate of the step size which is optimal in a quadratic potential.
After calculating the descent direction, a line search along this very direction is done to determine the -- with respect to the the Wolfe conditions -- optimal step size. 
This is iterated until the convergence criterion is satisfied or the maximum number of steps has been reached.
 
\subsection{Parallelization \& Cluster Compatibility}\label{sec:parallelization}
The software package \textsc{D2O} \citep{Steininger2016} was originally developed for massively parallelizing \nifty\ by distributing the fields' data arrays among multiple nodes in a MPI cluster. 
If two fields are distributed with the same distribution strategy individual nodes hold the same part of each of the fields' arrays.
With this ansatz, numerical operations like adding two fields involve no communication and hence exhibit excellent scaling behavior. 
Please refer to \citet{Steininger2016} for an exhaustive discussion.
As a consequence, in contrast to \niftyo\, the arrays in \niftyt\ are \verb|distributed_data_object|s, rather than \verb|numpy.ndarray|s.
Due to this strong encapsulation, most of its code base and the work with \niftyt\ is agnostic of parallelization. 
By utilizing \textsc{D2O}, \nifty\ can operate on high performance computing clusters and thus terabytes of RAM.
This renders a large quantity of new applications possible, like high-resolution runs of 3D reconstructions, cf.\ \citep{2016A&A...590A..59G,2017arXiv170101886B}, and Faraday synthesis algorithms \citep{2012A&A...540A..80B}.\\
Another side product of \niftyt\ is the \emph{keepers} package\footnote{\url{https://gitlab.mpcdf.mpg.de/ift/keepers}}, which is aimed at making scientific research on clusters as convenient as possible. 
To be precise, this involves classes and functionality for cluster-compatible logging, convenient algorithm parametrization, storing \nifty\ objects to disk in a versioned fashion and therefore allowing for restartable jobs. 

\section{Application: Wiener Filter Reconstructions}\label{sec:wiener_filter}
\subsection{Case 1: Single Space Geometry}\label{sec:}
\begin{figure*}
\begin{subfigure}{.5\textwidth}
  \centering
  \includegraphics[width=.98\linewidth]{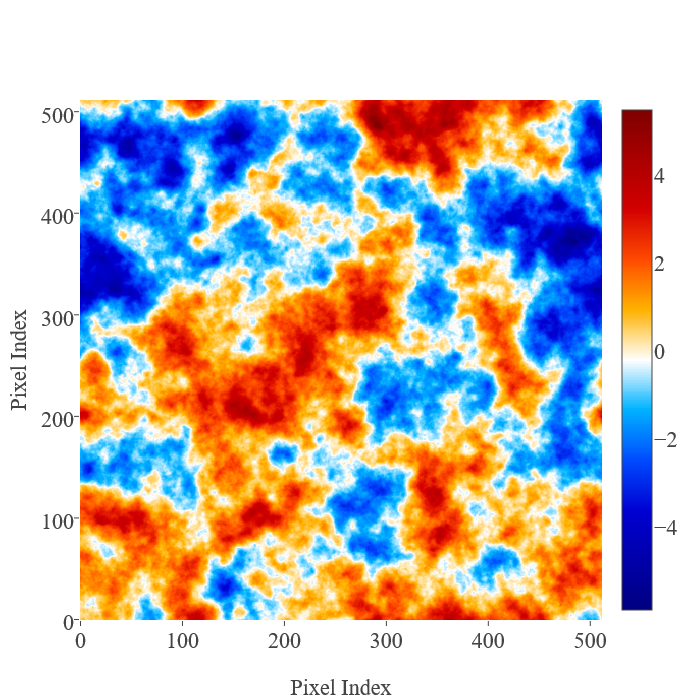}
  \caption{Mock signal.}
%  \label{fig:sfig1}
\end{subfigure}%
\begin{subfigure}{.5\textwidth}
  \centering
  \includegraphics[width=.98\linewidth]{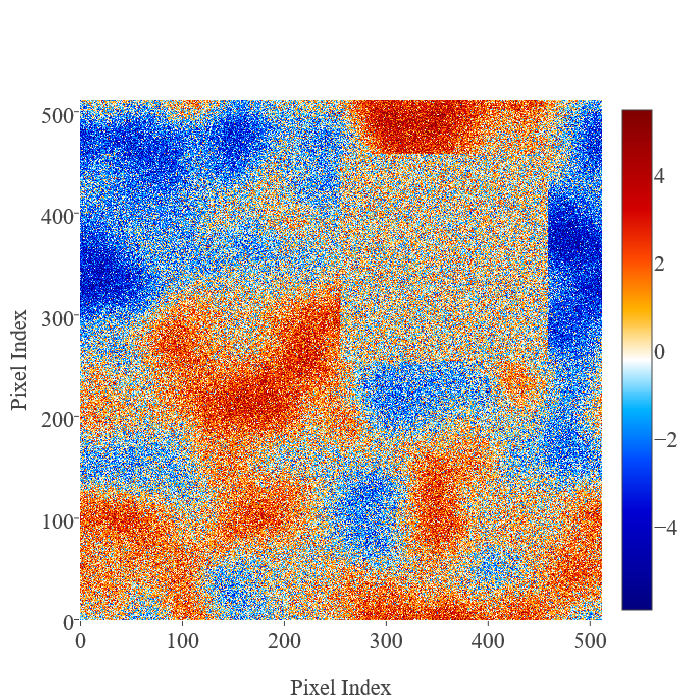}
  \caption{Data, which is the result of smoothing, masking and additive noise.}
%  \label{fig:sfig2}
\end{subfigure}\\
\begin{subfigure}{.5\textwidth}
  \centering
  \includegraphics[width=.98\linewidth]{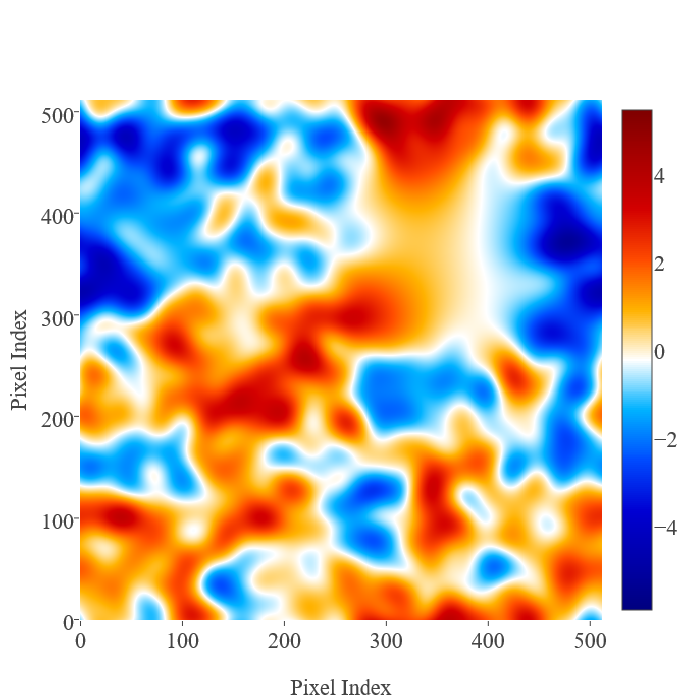}
  \caption{Maximum a-posteriori reconstruction.}
%  \label{fig:sfig2}
\end{subfigure}
\begin{subfigure}{.5\textwidth}
  \centering
  \includegraphics[width=.98\linewidth]{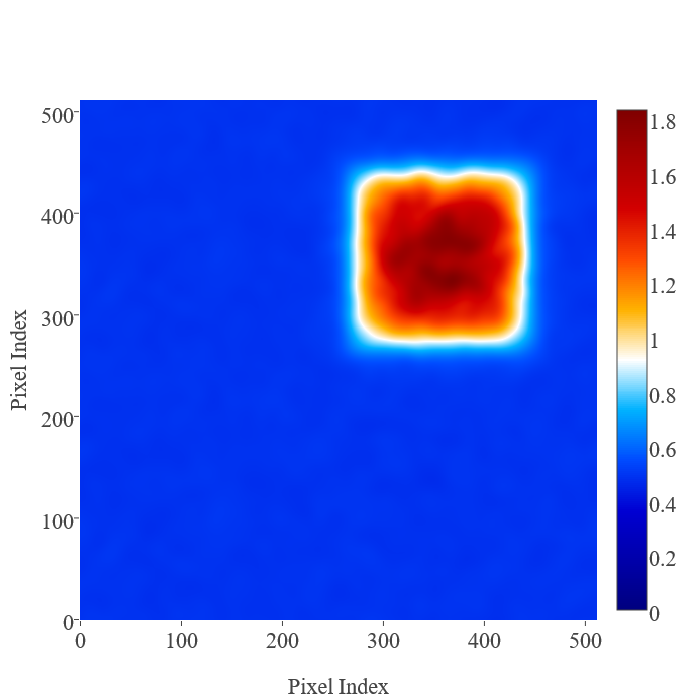}
  \caption{Uncertainty map.}
%  \label{fig:sfig2}
\end{subfigure}
\caption{Illustration of a Wiener filter reconstruction. 
The labeling of the axes shows pixel numbers.}
\label{fig:wiener_filter_plots}
\end{figure*}

Figure \ref{fig:wiener_filter_demo_code} shows an exemplary implementation of the Wiener filter with \niftyt. 
First, the parameters that characterize the mock signal's correlation structure are set up (lines \ref{code:wf_parameters}ff.). 
Then, starting with line \ref{code:wf_geometry}, the domain geometry is specified: here, a single regular gridded two-dimensional Cartesian space is used. 
Afterwards, in lines \ref{code:wf_mock_signal}ff., a signal covariance is defined to create the mock signal that shall be reconstructed by the Wiener filter later.
In line \ref{code:wf_response} an exemplary response operator is set up which acts via smoothing and masking on the input signal. 
After setting up a noise covariance and creating a noise sample, in line \ref{code:wf_mock_data} the mock data is created.
In line \ref{code:wf_wiener_filter} the actual Wiener filter is performed by applying the propagator operator $D$, which coincides with the inverse of the information Hamiltonian's curvature. 
Once the reconstruction is done, an uncertainty map is computed (lines \ref{code:wf_uncertainty_probing}ff.).
Therefore, a prober class with the desired probing-targets -- here only the diagonal -- is constructed via multiple inheritance.
After this, a prober instance is created, which is then applied to the Wiener filter curvature operator. 
Because the \verb|wiener_curvature| operates in harmonic space, but the probes must be evaluated in position space, Fourier transforms are wrapped around the operator in line \ref{code:wf_variance_fft_wrap}. 
Please note that the reconstruction's variance is given by the bare diagonal entries, viz.\ no volume factors included, of the inverse curvature, which is the reason for the inverse weighting in line \ref{code:wf_variance_weighting}. 
Finally, the lines \ref{code:wf_plotting}ff.\ produce the plots that are shown in figure \ref{fig:wiener_filter_plots}.

\begin{figure*}[!t]
    \begin{lstlisting}[language=iPython]
    import nifty as ift
    import numpy as np
    from keepers import Repository
    
    if __name__ == "__main__":
        ift.nifty_configuration['default_distribution_strategy'] = 'fftw'
    
        # Setting up parameters    |\label{code:wf_parameters}|
        correlation_length_scale = 1.  # Typical distance over which the field is correlated
        fluctuation_scale = 2.         # Variance of field in position space
        response_sigma = 0.05          # Smoothing length of response (in same unit as L)
        signal_to_noise = 1.5          # The signal to noise ratio
        np.random.seed(43)             # Fixing the random seed
        def power_spectrum(k):         # Defining the power spectrum
            a = 4 * correlation_length_scale * fluctuation_scale**2
            return a / (1 + (k * correlation_length_scale)**2) ** 2
    
        # Setting up the geometry |\label{code:wf_geometry}|
        L = 2.  # Total side-length of the domain
        N_pixels = 512  # Grid resolution (pixels per axis)
        signal_space = ift.RGSpace([N_pixels, N_pixels], distances=L/N_pixels)
        harmonic_space = ift.FFTOperator.get_default_codomain(signal_space)
        fft = ift.FFTOperator(harmonic_space, target=signal_space, target_dtype=np.float)
        power_space = ift.PowerSpace(harmonic_space)
    
        # Creating the mock signal |\label{code:wf_mock_signal}|
        S = ift.create_power_operator(harmonic_space, power_spectrum=power_spectrum)
        mock_power = ift.Field(power_space, val=power_spectrum)
        mock_signal = fft(mock_power.power_synthesize(real_signal=True))
    
        # Setting up an exemplary response
        mask = ift.Field(signal_space, val=1.)
        N10 = int(N_pixels/10)
        mask.val[N10*5:N10*9, N10*5:N10*9] = 0.
        R = ift.ResponseOperator(signal_space, sigma=(response_sigma,), exposure=(mask,))  |\label{code:wf_response}|
        data_domain = R.target[0]
        R_harmonic = ift.ComposedOperator([fft, R], default_spaces=[0, 0])
    
        # Setting up the noise covariance and drawing a random noise realization
        N = ift.DiagonalOperator(data_domain, diagonal=mock_signal.var()/signal_to_noise, 
                                 bare=True)
        noise = ift.Field.from_random(domain=data_domain, random_type='normal',
                                      std=mock_signal.std()/np.sqrt(signal_to_noise), mean=0)
        data = R(mock_signal) + noise  |\label{code:wf_mock_data}|
    
        # Wiener filter
        j = R_harmonic.adjoint_times(N.inverse_times(data))
        wiener_curvature = ift.library.WienerFilterCurvature(S=S, N=N, R=R_harmonic)
        m_k = wiener_curvature.inverse_times(j)  |\label{code:wf_wiener_filter}|
        m = fft(m_k)
    
        # Probing the uncertainty  |\label{code:wf_uncertainty_probing}|
        class Proby(ift.DiagonalProberMixin, ift.Prober): pass
        proby = Proby(signal_space, probe_count=800)
        proby(lambda z: fft(wiener_curvature.inverse_times(fft.inverse_times(z)))) |\label{code:wf_variance_fft_wrap}|
        sm = ift.SmoothingOperator(signal_space, sigma=0.03)
        variance = ift.sqrt(sm(proby.diagonal.weight(-1)))  |\label{code:wf_variance_weighting}| 
     
        # Plotting |\label{code:wf_plotting}|
        plotter = ift.plotting.RG2DPlotter(color_map=plotting.colormaps.PlankCmap())
        plotter.figure.xaxis = ift.plotting.Axis(label='Pixel Index')
        plotter.figure.yaxis = ift.plotting.Axis(label='Pixel Index')
        plotter.plot.zmax = variance.max(); plotter.plot.zmin = 0
        plotter(variance, path = 'uncertainty.html')
        plotter.plot.zmax = mock_signal.max(); plotter.plot.zmin = mock_signal.min()
        plotter(mock_signal, path='mock_signal.html')
        plotter(ift.Field(signal_space, val=data.val), path='data.html')
        plotter(m, path='map.html')
    \end{lstlisting}
\caption{A full-feature Wiener filter implementation in \niftyt, including mock-data creation, signal reconstruction, uncertainty estimation, and plotting of the results.}
\label{fig:wiener_filter_demo_code}
\end{figure*}

\subsection{Case 2: Cartesian Product Space Geometry}
\begin{figure*}
\begin{subfigure}{.5\textwidth}
  \centering
  \includegraphics[width=.98\linewidth]{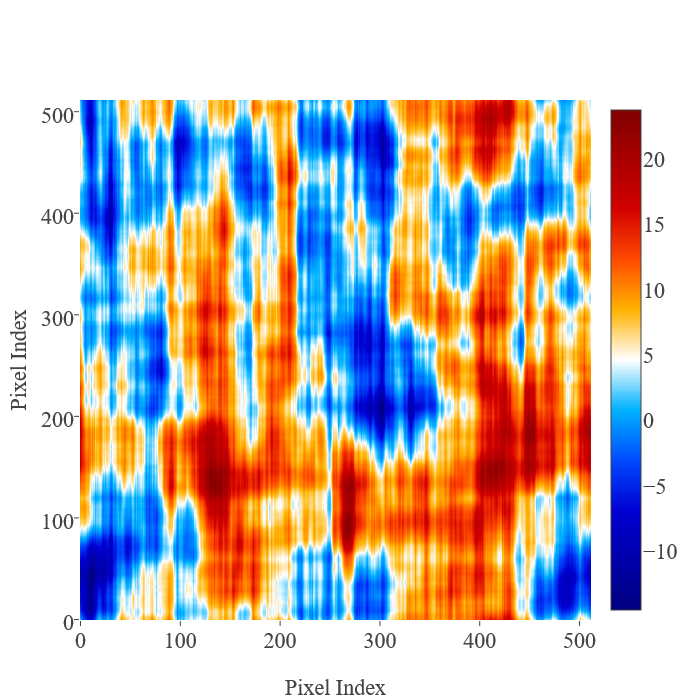}
  \caption{Mock signal with two individual power spectra for each space.}
%  \label{fig:sfig1}
\end{subfigure}%
\begin{subfigure}{.5\textwidth}
  \centering
  \includegraphics[width=.98\linewidth]{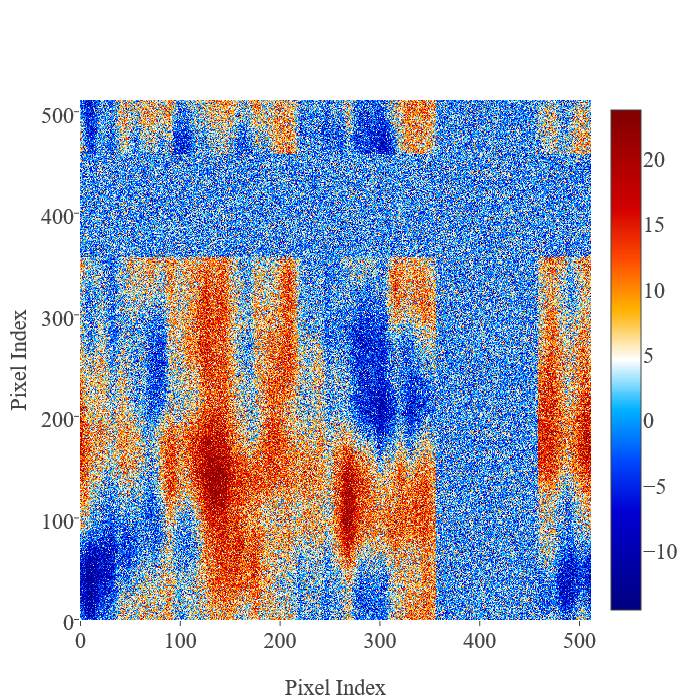}
  \caption{Data, which is the result of smoothing, masking and additive noise.}
%  \label{fig:sfig2}
\end{subfigure}\\
\begin{subfigure}{.5\textwidth}
  \centering
  \includegraphics[width=.98\linewidth]{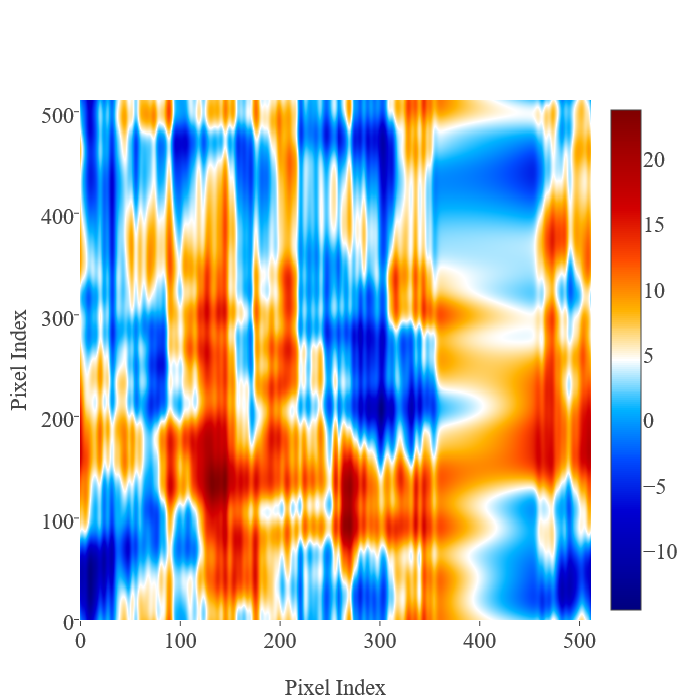}
  \caption{Maximum a-posteriori reconstruction.}
%  \label{fig:sfig2}
\end{subfigure}
\begin{subfigure}{.5\textwidth}
  \centering
  \includegraphics[width=.98\linewidth]{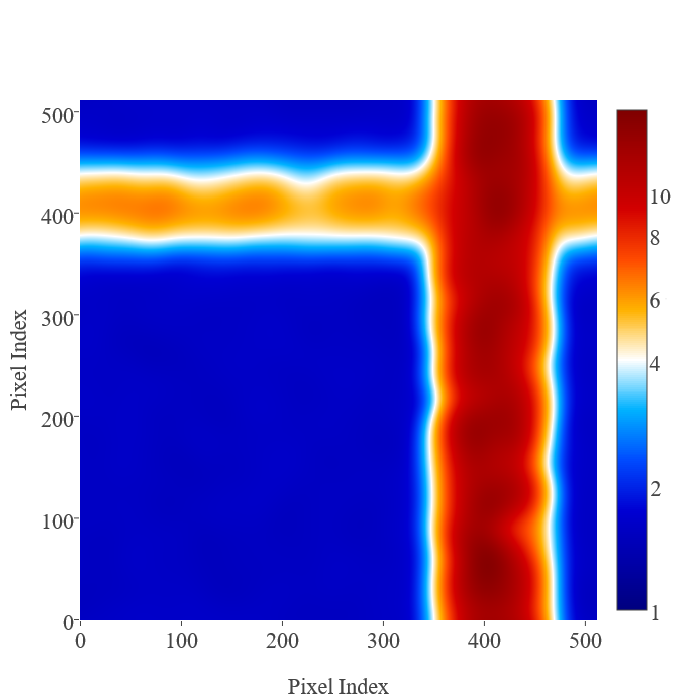}
  \caption{Uncertainty map shown on a logarithmic scale.}
%  \label{fig:sfig2}
\end{subfigure}
\caption{Illustration of a Wiener filter reconstruction in the context of the Cartesian product of two one-dimensional regular grid spaces. 
The labeling of the axes shows pixel numbers.}
\label{fig:cartesian_wiener_filter_plots}
\end{figure*}

As discussed in section \ref{sec:domain_objects}, one of the crucial features of \niftyt\ is that fields can be defined on the Cartesian product of multiple individual spaces. 
This makes it possible to easily implement inference algorithms that reconstruct signals which have a mixed correlation structure. 
In the above case, a Wiener filter was applied in the context of a two-dimensional regular grid geometry. 
With \niftyt\ this can easily be extended to, for example, the case of the Cartesian product of two one-dimensional regular grid spaces, shown in figure \ref{fig:cartesian_wiener_filter_plots}.
Now the power spectrum for each space-component of the signal field differs. 
Additionally, the response operator has individual smoothing lengths and masks for each of the spaces.
In this example, the power spectrum of the space drawn on the vertical axis is steeper than the one of the horizontal space. 
As a consequence, the field possesses small scale structure primarily in the horizontal direction. 
A remarkable consequence of this is reflected in the uncertainty map. 
Although the occlusion mask is equally broad for each space, the uncertainty is much higher in the direction of the horizontal space. 
There, the Wiener filter is not able to interpolate as well as for the large scale dominated vertical space. \\
Since the structure of the code for the case of a Cartesian product of spaces is very similar to the one in figure \ref{fig:wiener_filter_demo_code} -- actually, one mainly has to define the two individual geometries respectively -- the code is not shown explicitly here.  
For interested readers, the code that was used to produce the plots given in figure \ref{fig:cartesian_wiener_filter_plots} is available as a demo in the \niftyt\ code release, which is available here: \url{https://gitlab.mpcdf.mpg.de/ift/NIFTy}. 

\section{Conclusion}
\niftyt\ allows the programming of field equations independent of the underlying geometry or resolution. 
This freedom is particularly desirable in the implementation of inference algorithms of continuous quantities. 
This behavior is achieved by an object-oriented structure which cleanly separates the abstract mathematical operations from the underlying numerical calculations.
By using layers of abstraction the operations are kept general and simple while preserving the continuum limit. 
Normalizations are applied automatically without the need of specification by the user. 
\niftyt\ comes with full support of n-dimensional Cartesian spaces, the surface of the sphere and product spaces generated from them. 
Other geometries can be included in a straightforward fashion due to the layers of abstraction in \niftyt. \\
Algorithms implemented with \niftyt\ are suitable to be run on HPC clusters with (almost) no MPI awareness needed from the user. 
With \niftyt\ formulas can be transformed easily into code and, conversely, code can be easily read as formulas. 
This allows both, rapid prototyping and the implementation of large-scale algorithms. \\
\niftyt\ has been developed to ease the implementation of inference algorithms on continuous quantities. 
Therefore, it is applicable in various areas. 
The first version of \nifty\ already enabled numerous inference projects including many imaging algorithms performing interferometry, tomography (both astronomical and medical) or deconvolution, but also more abstract inference algorithms such as the estimation of cosmological parameters or instrument calibration.
With the additional power of \niftyt\ it can already be observed that this variety now grows even further. \\
\niftyt\ is open-source software available under the GNU General Public License v3 (GPL-3) at \url{https://gitlab.mpcdf.mpg.de/ift/NIFTy/}.

\begin{acknowledgements}
      Part of this work was supported by the \emph{Studienstiftung des deutschen Volkes}.
\end{acknowledgements}

\clearpage

\bibliographystyle{aa}
\bibliography{ift,NIFTY}

\appendix

\begin{landscape}
\section{Class Structure}
    \begin{figure}[h!]
        \centering
        \begin{tikzpicture}[>=stealth,thick, every node/.style={shape=rectangle,draw,rounded corners,align=left}]
            \node[rectangle split, rectangle split parts=3] (DomainObject) at (1, 5)
            {\textit{\texttt{DomainObject}}
                \nodepart[text width=5cm]{second}
                \begin{varwidth}{\textwidth}
                    \begin{itemize}[itemsep=0.25ex]
                        \item[] \texttt{*shape}: \texttt{tuple}
                        \item[] \texttt{*dim}: \texttt{int}
                    \end{itemize}
                \end{varwidth}
                \nodepart[text width=5cm]{third}
                \begin{varwidth}{\textwidth}
                    \begin{itemize}[itemsep=0.25ex]
                        \item[] \texttt{*weight(data, \dots)}
                        \item[] \texttt{pre\_cast(data, axes)}
                        \item[] \texttt{post\_cast(data, axes)}
                    \end{itemize}
                \end{varwidth}
            };
            \node[rectangle split, rectangle split parts=3] (Space) at (5, 0)
            {\textit{\texttt{Space}}
                \nodepart[text width=8.5cm]{second}
                \begin{varwidth}{\textwidth}
                    \begin{itemize}[itemsep=0.25ex]
                        \item[] \texttt{*harmonic}: \texttt{boolean}
                        \item[] \texttt{*total\_volume}: \texttt{int}
                    \end{itemize}
                \end{varwidth}
                \nodepart[text width=8.5cm]{third}
                \begin{varwidth}{\textwidth}
                    \begin{itemize}[itemsep=0.25ex]
                        \item[] \texttt{*copy()}
                        \item[] \texttt{*hermitian\_decomposition(data, \dots)}
                        \item[] \texttt{*get\_distance\_array(distribution\_strategy)}
                        \item[] \texttt{*get\_fft\_smoothing\_kernel\_function(sigma)}
                    \end{itemize}
                \end{varwidth}
            };
            \node[rectangle split, rectangle split parts=2] (RGSpace) at (0, -4)
            {\texttt{RGSpace}
                \nodepart[text width=4cm]{second}
                \begin{varwidth}{\textwidth}
                    \begin{itemize}[itemsep=0.25ex]
                        \item[] \texttt{zerocenter}: \texttt{tuple}
                        \item[] \texttt{distances}: \texttt{tuple}
                    \end{itemize}
                \end{varwidth}
            };
            \node[rectangle split, rectangle split parts=2] (HPSpace) at (5, -4)
            {\texttt{HPSpace}
                \nodepart[text width=3cm]{second}
                \begin{varwidth}{\textwidth}
                    \begin{itemize}[itemsep=0.25ex]
                        \item[] \texttt{nside}: \texttt{int}
                    \end{itemize}
                \end{varwidth}
            };
            \node[rectangle split, rectangle split parts=2] (GLSpace) at (10, -4)
            {\texttt{GLSpace}
                \nodepart[text width=3cm]{second}
                \begin{varwidth}{\textwidth}
                    \begin{itemize}[itemsep=0.25ex]
                        \item[] \texttt{nlat}: \texttt{int}
                        \item[] \texttt{nlon}: \texttt{int}
                    \end{itemize}
                \end{varwidth}
            };
            \node[rectangle split, rectangle split parts=2] (LMSpace) at (0, -7)
            {\texttt{LMSpace}
                \nodepart[text width=2.5cm]{second}
                \begin{varwidth}{\textwidth}
                    \begin{itemize}[itemsep=0.25ex]
                        \item[] \texttt{lmax}: \texttt{int}
                        \item[] \texttt{mmax}: \texttt{int}
                    \end{itemize}
                \end{varwidth}
            };
            \node[rectangle split, rectangle split parts=2] (PowerSpace) at (7.5, -7)
            {\texttt{PowerSpace}
                \nodepart[text width=7.8cm]{second}
                \begin{varwidth}{\textwidth}
                    \begin{itemize}[itemsep=0.25ex]
                        \item[] \texttt{config}: \texttt{dict}
                        \item[] \texttt{harmonic\_partner}: \texttt{Space}
                        \item[] \texttt{kindex, pundex, rho}: \texttt{np.ndarray}
                        \item[] \texttt{pindex, k\_array}: \texttt{distributed\_data\_object}
%                        \item[] \texttt{pindex}: \texttt{distributed\_data\_object}
%                        \item[] \texttt{pundex}: \texttt{np.ndarray}
%                        \item[] \texttt{rho}: \texttt{np.ndarray}
                    \end{itemize}
                \end{varwidth}
            };
            \node[rectangle split, rectangle split parts=2] (FieldType) at (-3, 1)
            {\texttt{FieldType}
                \nodepart[text width=4cm]{second}
                \begin{varwidth}{\textwidth}
                    \begin{itemize}[itemsep=0.25ex]
                        \item[] \texttt{weight(data, \dots)}
                    \end{itemize}
                \end{varwidth}
            };
            \node[rectangle split, rectangle split parts=1] (FieldArray) at (-3, -1)
            {\texttt{FieldArray}
%                \nodepart[text width=4cm]{second}
%                \begin{varwidth}{\textwidth}
%                    \begin{itemize}[itemsep=0.25ex]
%                        \item[] \texttt{shape: tuple}
%                        \item[] \texttt{dim: int}
%                    \end{itemize}
%                \end{varwidth}
            };
            \draw [-open triangle 45] (FieldArray.north) -- (FieldType.south);
            \draw [-open triangle 45] (FieldType.north) |- ($(DomainObject.south)-(0,0.75)$) -| (DomainObject.south);
            \draw [-open triangle 45] (Space.north) |- ($(DomainObject.south)-(0,0.75)$) -| (DomainObject.south);
            \draw [-open triangle 45] (RGSpace.north) |- ($(Space.south)-(0,0.75)$) -| (Space.south);
            \draw [-open triangle 45] (HPSpace.north) -- (Space.south);
            \draw [-open triangle 45] (GLSpace.north) |- ($(Space.south)-(0,0.75)$) -| (Space.south);
            \draw [-open triangle 45] (LMSpace.north) |- ($(LMSpace.north)+(0,0.5)$) -- ($(LMSpace.east)+(1.45, 1.3)$) |- ($(Space.south)-(0,0.75)$) -| (Space.south);
            \draw [-open triangle 45] (PowerSpace.north) |- ($(Space.south)-(0,0.75)$) -| (Space.south);
        \end{tikzpicture}
        \caption{UML diagram that shows \texttt{DomainObject} and its descendants. 
                 For the derived classes only added attributes and methods are given. 
                 All superordinate attributes and methods apply implicitly to the child classes, too, since \niftyt\ strictly obeys the Liskov substitution principle.
                 Abstract attributes and methods are denoted with a \texttt{*}; 
                 Unless specified explicitly, they are implemented only by the leafs of the inheritance tree. 
                 }
        \label{fig:DomainObjects}
    \end{figure}
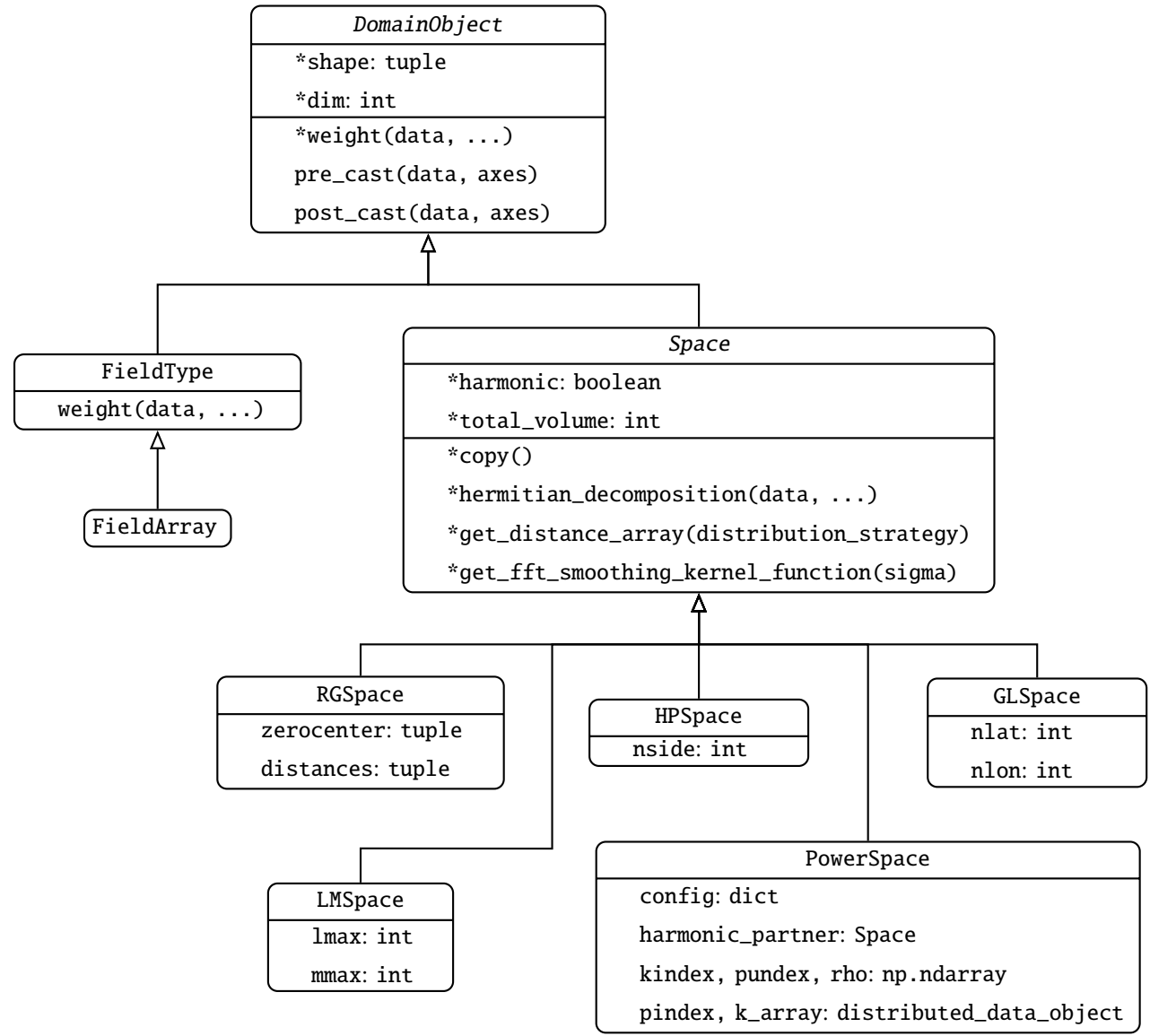
\end{landscape}

\begin{landscape}
    \begin{figure}[h!]
        \centering
        \begin{tikzpicture}[>=stealth,thick, every node/.style={shape=rectangle,draw,rounded corners,align=left}, anchor=north]
            \node[rectangle split, rectangle split parts=3] (LinearOperator) at (-5, 6.5)
            {\textit{\texttt{LinearOperator}}
                \nodepart[text width=7cm]{second}
                \begin{varwidth}{\textwidth}
                    \begin{itemize}[itemsep=0.25ex]
                        \item[] \texttt{*domain}: \texttt{tuple of DomainObjects}
                        \item[] \texttt{*target}: \texttt{tuple of DomainObjects}
                        \item[] \texttt{*unitary}: \texttt{bool}
                    \end{itemize}
                \end{varwidth}
                \nodepart[text width=7cm]{third}
                \begin{varwidth}{\textwidth}
                    \begin{itemize}[itemsep=0.25ex]
                        \item[] $\dagger$\texttt{times(data, \dots)}
                        \item[] $\dagger$\texttt{inverse\_times(data, \dots)}
                        \item[] $\dagger$\texttt{adjoint\_times(data, \dots)}
                        \item[] $\dagger$\texttt{adjoint\_inverse\_times(data, \dots)}
                        \item[] $\dagger$\texttt{inverse\_adjoint\_times(data, \dots)}
                    \end{itemize}
                \end{varwidth}
            };
            \node[rectangle split, rectangle split parts=3] (FFTOperator) at (-15, 0)
            {\texttt{FFTOperator}
                \nodepart{second}
                \nodepart[text width=5cm]{third}
                \begin{varwidth}{\textwidth}
                    \begin{itemize}[itemsep=0.25ex]
                        \item[] \texttt{times(data, \dots)}
                        \item[] \texttt{adjoint\_times(data, \dots)}
                    \end{itemize}
                \end{varwidth}
            };
            \node[rectangle split, rectangle split parts=3] (ComposedOperator) at (-9, 0)
            {\texttt{ComposedOperator}
                \nodepart{second}
                \nodepart[text width=6cm]{third}
                \begin{varwidth}{\textwidth}
                    \begin{itemize}[itemsep=0.25ex]
                        \item[] \texttt{times(data, \dots)}
                        \item[] \dots
                        \item[] \texttt{inverse\_adjoint\_times(\dots)}
                    \end{itemize}
                \end{varwidth}
            };
            \node[rectangle split, rectangle split parts=3] (ResponseOperator) at (-3, 0)
            {\texttt{ResponseOperator}
                \nodepart{second}
                \nodepart[text width=5cm]{third}
                \begin{varwidth}{\textwidth}
                    \begin{itemize}[itemsep=0.25ex]
                        \item[] \texttt{times(data, \dots)}
                        \item[] \texttt{adjoint\_times(data, \dots)}
                    \end{itemize}
                \end{varwidth}
            };
            \node[rectangle split, rectangle split parts=3] (EndomorphicOperator) at (3.5, 0)
            {\textit{\texttt{EndomorphicOperator}}
                \nodepart[text width=7cm]{second}
                \begin{varwidth}{\textwidth}
                    \begin{itemize}[itemsep=0.25ex]
                        \item[] \texttt{target := domain} 
                        \item[] \texttt{*self\_adjoint}: \texttt{tuple}
                    \end{itemize}
                \end{varwidth}
                \nodepart[text width=7cm]{third}
                \begin{varwidth}{\textwidth}
                    \begin{itemize}[itemsep=0.25ex]
                        \item[] \texttt{inverse\_times(data, \dots)}
                        \item[] \texttt{adjoint\_times(data, \dots)}
                        \item[] \texttt{adjoint\_inverse\_times(data, \dots)}
                        \item[] \texttt{inverse\_adjoint\_times(data, \dots)}
                    \end{itemize}
                \end{varwidth}                
            };
            \node[rectangle split, rectangle split parts=3] (SmoothingOperator) at (-15, -6.5)
            {\textit{\texttt{SmoothingOperator}}
                \nodepart[text width=5cm]{second}
                \begin{varwidth}{\textwidth}
                    \begin{itemize}[itemsep=0.25ex]
                        \item[] \texttt{sigma}: \texttt{float}
                        \item[] \texttt{log\_distances}: \texttt{bool}
                    \end{itemize}
                \end{varwidth}
                \nodepart[text width=5cm]{third}
                \begin{varwidth}{\textwidth}
                \begin{itemize}[itemsep=0.25ex]
                    \item[] \texttt{times(data, \dots)}
                    \item[] \texttt{inverse\_times(data, \dots)}
                \end{itemize}
                \end{varwidth}    
            };

            \node[rectangle split, rectangle split parts=3] (SmoothnessOperator) at (-9, -6.5)
            {\textit{\texttt{SmoothnessOperator}}
                \nodepart[text width=5cm]{second}
                \begin{varwidth}{\textwidth}
                    \begin{itemize}[itemsep=0.25ex]
                        \item[] \texttt{strength}: \texttt{float}
                        \item[] \texttt{logarithmic}: \texttt{bool}
                    \end{itemize}
                \end{varwidth}
                \nodepart[text width=5cm]{third}
                \begin{varwidth}{\textwidth}
                \begin{itemize}[itemsep=0.25ex]
                    \item[] \texttt{times(data, \dots)}
                \end{itemize}
                \end{varwidth}    
            };
            
            \node[rectangle split, rectangle split parts=3] (LaplaceOperator) at (-15, -3)
            {\textit{\texttt{LaplaceOperator}}
                \nodepart[text width=5cm]{second}
                \nodepart[text width=5cm]{third}
                \begin{varwidth}{\textwidth}
                \begin{itemize}[itemsep=0.25ex]
                    \item[] \texttt{times(data, \dots)}                    
                    \item[] \texttt{adjoint\_times(data, \dots)}
                \end{itemize}
                \end{varwidth}    
            };            

            \node[rectangle split, rectangle split parts=3] (ProjectionOperator) at (-3.25, -6.5)
            {\texttt{ProjectionOperator}
                \nodepart[text width=5cm]{second}
                \begin{varwidth}{\textwidth}
                    \begin{itemize}[itemsep=0.25ex]
                        \item[] \texttt{projection\_field}: \texttt{Field}
                    \end{itemize}
                \end{varwidth}
                \nodepart[text width=5cm]{third}
                \begin{varwidth}{\textwidth}
                    \begin{itemize}[itemsep=0.25ex]
                        \item[] \texttt{times(data, \dots)}
                        \item[] \texttt{inverse\_times(data, \dots)}
                    \end{itemize}
                \end{varwidth}  
            };
            \node[rectangle split, rectangle split parts=3] (DiagonalOperator) at (3.5, -6.5)
            {\texttt{DiagonalOperator}
                \nodepart[text width=7cm]{second}
                \begin{varwidth}{\textwidth}
                    \begin{itemize}[itemsep=0.25ex]
                        \item[] \texttt{diagonal}: \texttt{Field}
                    \end{itemize}
                \end{varwidth}
                
                \nodepart[text width=7cm]{third}
                \begin{varwidth}{\textwidth}
                    \begin{itemize}[itemsep=0.25ex]
                        \item[] \texttt{times(data, \dots)}
                        \item[] \texttt{inverse\_times(data, \dots)}
                        \item[] \texttt{adjoint\_times(data, \dots)}
                        \item[] \texttt{adjoint\_inverse\_times(data, \dots)}
                    \end{itemize}
                \end{varwidth} 
            };
            \draw [-open triangle 45] (FFTOperator.north) |- ($(LinearOperator.south)-(0,0.75)$) -| (LinearOperator.south);
            \draw [-open triangle 45] (ComposedOperator.north) |- ($(LinearOperator.south)-(0,0.75)$) -| (LinearOperator.south);
            \draw [-open triangle 45] (ResponseOperator.north) |- ($(LinearOperator.south)-(0,0.75)$) -| (LinearOperator.south);
            \draw [-open triangle 45] (EndomorphicOperator.north) |- ($(LinearOperator.south)-(0,0.75)$) -| (LinearOperator.south);
            \draw [-open triangle 45] (SmoothingOperator.north) |- ($(EndomorphicOperator.south)-(0,1.75)$) -| (EndomorphicOperator.south); 
            \draw [-open triangle 45] (LaplaceOperator.south) |- ($(EndomorphicOperator.south)-(0,1.75)$) -| (EndomorphicOperator.south); 
            \draw [-open triangle 45] (SmoothnessOperator.north) |- ($(EndomorphicOperator.south)-(0,1.75)$) -| (EndomorphicOperator.south);
            \draw [-open triangle 45] (ProjectionOperator.north) |- ($(EndomorphicOperator.south)-(0,1.75)$) -| (EndomorphicOperator.south);
            \draw [-open triangle 45] (DiagonalOperator.north) |- ($(EndomorphicOperator.south)-(0,1.75)$) -| (EndomorphicOperator.south);
        \end{tikzpicture}
        \caption{UML diagram that shows the inheritance structure for the \nifty\ \texttt{Operator}s.
            For the derived classes only added attributes and methods are given. 
            All superordinate attributes and methods apply implicitly to the child classes, too, since \niftyt\ strictly obeys the Liskov substitution principle.
            Abstract attributes and methods are denoted with a \texttt{*}; 
            Unless specified explicitly, they are implemented only by the leafs of the inheritance tree.            
            For the \texttt{LinearOperator} class, the $\dagger$ signals, that the methods are defined, though not implemented.}
        \label{fig:Operators}
            
    \end{figure}
\end{landscape}

\end{document}